\documentclass[USenglish,oneside,twocolumn]{article}

\usepackage[utf8]{inputenc}
\usepackage[big]{dgruyter_NEW}

\usepackage{balance}

\usepackage{natbib}
\usepackage{amsmath,amssymb,amsfonts}
\usepackage{algorithmic}
\usepackage{graphicx}
\usepackage{textcomp}
\usepackage{todonotes}
\usepackage{amsmath,amsfonts}
\usepackage{algorithmic}
\usepackage{graphicx}
\usepackage{url}
\usepackage{textcomp}
\usepackage{graphicx,balance,multirow,xspace,subcaption}
\usepackage{multirow,tabularx}
\usepackage{pifont}
\usepackage{todonotes}
\usepackage{pifont}
\PassOptionsToPackage{hyphens}{url}\usepackage{hyperref}
\usepackage{graphicx,balance,multirow,xspace,subcaption}
\usepackage{xcolor}
\usepackage{multirow}
\usepackage[draft,inline,nomargin,index]{fixme}
\usepackage{xspace}
\usepackage{todonotes}
\usepackage{paralist}
\usepackage{url}
\usepackage[normalem]{ulem}
\usepackage{float}
\usepackage{adjustbox}

\newcommand{\blocker}{\emph{IoTrimmer}\xspace}
\newcommand{\trigger}{\emph{IoTrigger}\xspace}
\newcommand{\iolist}{\emph{IoTrim}\xspace}
\newcommand{\allow}{\emph{allow-listing}\xspace}
\newcommand{\deny}{\emph{deny-listing}\xspace}

\def\etal{\emph{et al.\xspace}}
\def\ie{\emph{i.e.},\xspace}
\def\eg{\emph{e.g.},\xspace}
\def\etc{\emph{etc.}\xspace}

\newcommand\new[1]{#1}

\let\todonotes\todo
\renewcommand{\todo}[1]{\todonotes[inline,color=red]{TODO - #1}}

\newcommand{\third}{$3^{rd}$}

\newcommand{\numdev}{31\xspace}

\newcommand{\numdevblockable}{16\xspace}

\newcommand{\numdevnoblockable}{15\xspace}

\newcommand{\numreqdestnoblock}{22\xspace}

\newcommand{\numdevcommon}{three\xspace}

\newcommand{\numdest}{119\xspace}

\newcommand{\numdestnonreq}{62\xspace}

\newcommand{\maxdestnonreq}{11\xspace}

\newcommand{\trafficblocked}{1,931 bytes\xspace}

\newcommand{\numcategories}{five\xspace}

\newcommand{\nummonths}{six\xspace}

\newcommand{\numevaldays}{7\xspace}

\newcommand{\numfunct}{217\xspace}

\newenvironment{packed_itemize}{
\begin{itemize}
  \nointerlineskip
  \setlength{\itemsep}{0pt}
  \setlength{\parskip}{0pt}
  \setlength{\parsep}{0pt}
  \setlength{\topsep}{0pt}
  
}{\end{itemize}}

\usepackage[absolute,showboxes]{textpos}

\usepackage{blindtext,graphicx}
\usepackage[absolute]{textpos}
\newcommand{\arxiv}[1]{#1}

\begin{document}

\TPMargin{3pt}
\arxiv{\begin{textblock}{17}(4.35,3)
  \noindent
  \footnotesize
If you cite this paper, please use the PETS reference: Anna Maria Mandalari, Daniel J. Dubois, Roman Kolcun, Muhammad Talha Paracha, Hamed Haddadi, David Choffnes. Blocking without Breaking: Identification and Mitigation of Non-Essential IoT Traffic. In Privacy Enhancing Technologies Symposium (PETS) 2021.
\end{textblock}}

  \author*[1]{Anna Maria Mandalari}
  
 \author[2]{Daniel J. Dubois}

  \author[3]{Roman Kolcun}

  \author[4]{Muhammad Talha Paracha}
  
    \author[5]{Hamed Haddadi}

  \author[6]{David Choffnes}
  
    \affil[1]{Imperial College London, E-mail: anna-maria.mandalari@imperial.ac.uk}
  
    \affil[2]{Northeastern University, E-mail: d.dubois@northeastern.edu}
    
      \affil[3]{Imperial College London, E-mail: roman.kolcun@imperial.ac.uk}

  \affil[4]{Northeastern University, E-mail: paracha.m@husky.neu.edu}
  
    \affil[5]{Imperial College London, E-mail: h.haddadi@imperial.ac.uk}

  \affil[6]{Northeastern University, E-mail: choffnes@ccs.neu.edu}

\title{\huge Blocking without Breaking: Identification and Mitigation of Non-Essential IoT Traffic
}

  \runningtitle{Blocking without Breaking: Identification and Mitigation of Non-Essential IoT Traffic}

 \begin{abstract}
{

Despite the prevalence of Internet of Things (IoT) devices, there is little information about the purpose and risks of the Internet traffic these devices generate, and consumers have limited options for controlling those risks. 
A key open question is whether one can mitigate these risks by automatically blocking some of the Internet connections from IoT devices, without rendering the devices inoperable.

In this paper, we address this question by developing a rigorous methodology that relies on \new{automated} IoT-device experimentation to reveal which network connections (and the information they expose) are essential, and which are not. We further develop strategies to \emph{automatically} classify network traffic destinations as either required (\ie their traffic is \emph{essential} for devices to work properly) 
or not, hence allowing firewall rules to block traffic sent to non-required destinations without breaking the functionality of the device. We find that indeed \numdevblockable{} among the \numdev{} devices we tested have at least one blockable non-required destination, with the maximum number of blockable destinations for a device being \maxdestnonreq{}. We further analyze the destination of network traffic and find that all third parties observed in our experiments are blockable, while first and support parties are neither uniformly required or non-required. Finally, we demonstrate the limitations of existing blocklists on IoT traffic, propose a set of guidelines for automatically limiting non-essential IoT traffic, and we develop a prototype system that implements these guidelines.

}
\end{abstract}




\maketitle

\section{Introduction}
\label{sec:introduction}

Consumer Internet of Things (IoT) devices (\eg smart TVs, speakers, surveillance cameras, appliances, \etc) are rapidly gaining presence in homes, offices, and public spaces~\cite{IoT-stats}. 
While these devices often come with convenient services, they open the door to numerous privacy and security risks~\cite{moniotr, 10.1145/3319535.3354198, varmarken2020tv}.
These devices often expose information to a large number of destinations~\cite{moniotr, saidi2020haystack}, including third party advertising and tracking services.

A fundamental approach for mitigating such risks would be to automatically block any connections that are not essential for the essential functionality of a device. 
For this approach to work, we need a systematic approach to identify and block traffic that is not essential for a device to work, with little-to-no user configuration, and without causing any device malfunction.
Unfortunately, existing solutions are not sufficient for this purpose. 
Approaches such as Pi-hole~\cite{Pi-hole} block DNS requests for advertising and tracking services using blocklists, but destinations on those blocklists are often based on web tracking, thus missing blockable destinations for our IoT devices. 
While standard IoT security solutions might be able to arbitrarily block connections, they are unable to determine the consequences of any blocking on device functionality.


In this paper, we design and validate a methodology for automatically determining the necessity of the destinations contacted by an IoT device for the correct execution of its primary functionality.   
The intuition behind our approach is that IoT device functions can be invoked using interfaces amenable to automation (\eg using a voice synthesizer or scripting companion app interactions).
Further, one can automatically determine whether the execution of such functions has been successful, by observing the IoT device signals (\eg the screenshot of a companion app, or the network traffic patterns generated).
Based on these intuitions, our methodology can be used for a target device and selected functionality, to build a list of non-required destinations that can be automatically blocked, without breaking such functionality.
Similarly, we can build a list of required destinations that can be automatically allowed (to preserve functionality), while blocking the rest of the traffic.

The key building blocks of our system are: automatically interacting with devices to exercise their functions; systematically blocking one or more observed connections; and automatically determining whether each interaction was successful after blocking a connection. 
We use an extensive testbed and large number of trials to find that 28 out of~\numdev{} devices (across \numcategories{} categories) are amenable to fully automated blocking analysis.    

We then turn to analyzing the blockable destinations. 
We find that~\numdevblockable{} of our~\numdev{} devices contact at least one non-required destination (and as many as \maxdestnonreq{} destinations) to execute \new{their main functions}. 
Across all devices, we find that \numdestnonreq{} non-required destinations are contacted. 
We further analyze the destinations of network traffic and find that all third parties observed in our experiments are blockable, while first and support parties are neither uniformly required or non-required. 
Additionally, we show that uniformly blocking all \numdestnonreq{} non-required destinations for all devices can lead to breaking device functionality: \numdevcommon{} devices exhibit a required destination that is a non-required destination for a different device.
We find that non-required/required destinations do not change over time \new{for all the devices, and that, for 90.32\% of the devices, such destinations tend to be the same across different device functions}.
Finally, we propose a set of guidelines for automatically limiting non-essential IoT traffic, and we develop a prototype system that implements these guidelines. 

\noindent To summarize, our key contributions include:
\begin {packed_itemize} 
\item A methodology for determining required and non-required destinations by automatically executing IoT device functions and determining the execution outcome while blocking selected destinations.
\item An analysis of required/non-required destinations contacted by a diverse set of consumer IoT devices. 
\item \new{The design of a testing system (\trigger{}) and a blocking system (\blocker{}) that use our method for building the required and non-required destinations list (\iolist{} list), and use it to block non-essential traffic. \trigger, \blocker, and the \iolist{} list are publicly available at  \url{http://iotrim.net/}.}
\end {packed_itemize}


\section{Assumptions and Goals}
\label{sec:assumptions}

In this section, we set the assumptions, the definitions, the goals, and the non-goals for this work.

\subsection{Assumptions and Definitions}

\noindent \textbf{Threat Model.}
We assume a system composed of three entities: (\emph{i}) an off-the-shelf \emph{IoT device}, with the ability to communicate to any destinations over the Internet; (\emph{ii}) the \emph{network traffic destinations}, which include any Internet destinations that the IoT device creates a connection to; and (\emph{iii}) the \emph{user}, who has access to the IoT device functionality. 
In our threat model, we consider an IoT device and the network traffic destinations as potential adversaries, since the IoT device can potentially expose information about its users (the victim) to any destination. 
Since most IoT traffic is encrypted or encoded~\cite{moniotr} and the vast majority of IoT systems are closed, it is infeasible to perfectly infer what information is exposed through network connections using blackbox techniques.
Instead, we question whether a given connection is necessary for supporting a device's function (\eg ringing a doorbell), and if not, we consider the connection to be a threat for unnecessary data exposure, \new{in line with GDPR data minimization~\cite{GDPR} and purpose limitation principles~\cite{GDPRpurpose}.
Hence, blocking such traffic can potentially reduce information exposure for users without affecting the device functionality.}

%

\noindent \textbf{Essential Traffic Definition.}
We define \emph{essential traffic}, with respect to a given IoT device function, the network traffic that is essential to fulfill such function. 

\noindent \textbf{Required Destination Definition.}
We define as \emph{required} all the network traffic destinations that are contacted as part of essential traffic. 

\subsection{Goals}
\label{subsec:researchqs}
\begin{figure}[t]
  \centering
  \includegraphics[width=0.9\linewidth]{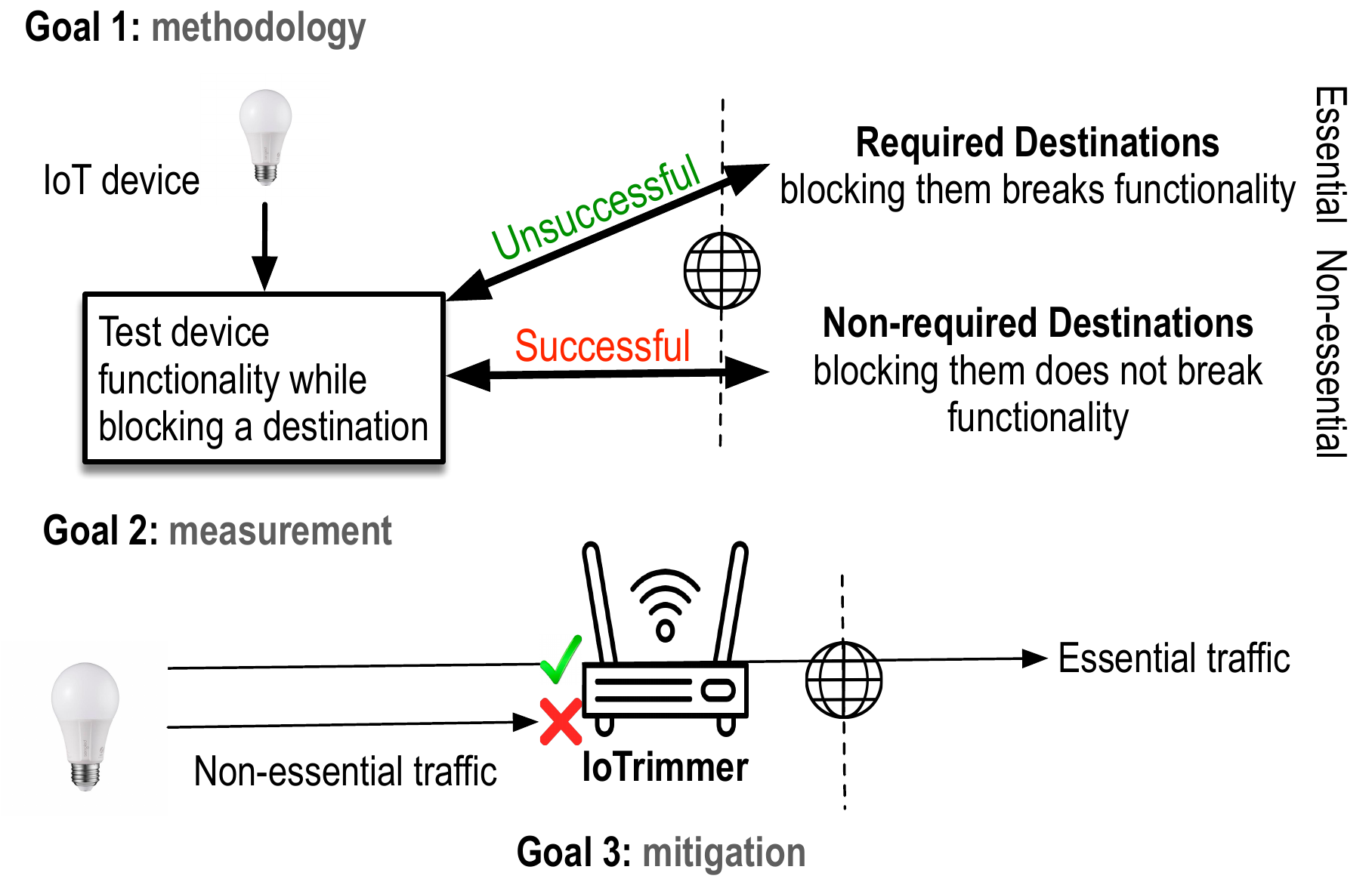}
  \caption{The three main goals of this paper.}
  \label{fig:goal}
\end{figure}

Fig.~\ref{fig:goal} illustrates the three main goals of this work.
More specifically, we want to answer the following questions.

\noindent \textbf{RQ1. How can we automatically identify non-essential IoT traffic?}
We seek to understand which destinations are not required for device functionality, so that we can block them to mitigate their potential risks. 
To address this, we propose a methodology for automatically detecting whether a network traffic destination is required or not for a given function of an IoT device (\eg in the case of a smart bulb and its switch on/off function, destinations that are not necessary for switching the light on/off).
The presence of a non-required destination means that all the traffic sent to such destination is non-essential, and therefore an avoidable case of information exposure.

\noindent \textbf{RQ2. What is the nature of non-essential IoT traffic?}
Armed with a measurement methodology to detect non-essential traffic, we apply it to identify and study the non-essential traffic produced by our set of \numdev{} popular IoT devices spanning \numcategories{} categories.
As part of this research question, we are interested in the type of destinations contacted (\eg if they belong to the device vendor), if any required destinations for a device are non-required for another device, if different devices have non-required destinations in common,
\new{if different device functions have different required and non-required destinations,}
\new{and if any of those destinations are present in existing blocklists.}
Characterizing non-essential traffic for existing devices is important to find correlations that can assist in detecting such traffic in real-time for future devices, without relying on the methodology proposed to answer RQ1.

\noindent \textbf{RQ3. Can we automatically mitigate non-essential IoT traffic?}
The knowledge of what destinations can be blocked for every device allows us to make automatic run-time decisions on what traffic to allow or not in a typical IoT deployment.
To answer this research question, we first determine the feasibility of a blocking solution by analyzing how much essential and non-essential traffic changes over time, as a way to assess the risk of allowing non-essential traffic or breaking the device functionality.
\new{Then, we describe our prototype software for automatically generating destination lists, and to transparently block as much non-essential IoT traffic as possible, thus reducing information exposure without affecting devices' main functions.}

\subsection{Non-Goals}
\label{subsec:nongoals}

In this work, we do not consider the following as goals, and leave them for future work.

\noindent \textbf{No control over how an IoT device works internally.} 
We consider the IoT devices as off-the-shelf consumer items that provide a finite set of functions and that communicate over the Internet.
For these devices, we have no control over their internal functions, but we can still interact with them using their user interface and we can measure their network activity.

\noindent \textbf{No content interception and inference.}
While we consider the visibility of the content out of scope, we are able to see the destinations of such traffic.
We make this assumption because the vast majority of the traffic is encrypted and the devices are assumed as blackboxes, where there is no possibility to install custom self-signed certificates to use man-in-the-middle techniques to intercept encrypted traffic.
We also do not try to infer the content of encrypted flows as means to measure privacy exposure since this has been studied in previous works, but we still use traffic patterns to test if blocking/allowing a destination prevents a given function from working.

\noindent \textbf{We do not test \emph{all} \new{functions}.}
IoT devices \new{typically offer several functions}; however, for this work, we apply our methodology by selecting only \new{a subset of them} for every IoT device under test so that we can have more coverage by devices rather than by functionality.
\new{We consider this limitation reasonable since our analysis of multiple functions in \S\ref{sub:add_funct} shows that the vast majority of the devices we tested use the same destinations for different functions.}
In this work, we assert that a given function is either executed correctly or not. We do not consider the case of a function partially working. 


\noindent \new{\textbf{One trigger per function.} Some IoT device functions can be triggered in several ways (\eg through a companion app, IFTTT~\cite{ifttt}, Samsung hub~\cite{smartthings}, \etc). In this work we only focus on one trigger per function.}


\section{Methodology}\label{sec:method}

We answer our first research question by proposing a methodology \new{to detect} non-essential IoT network traffic by classifying destinations as either required or not.

\subsection{Testbed}
\label{sub:testbed}

Our classification method relies on a testbed that provides a controlled environment for testing IoT devices.
Our testbed consists of: (\emph{i}) a \emph{router} that offers IP connectivity to the IoT devices under test, and the ability to capture and control network traffic for each device; and (\emph{ii}) a set of \emph{support scripts} to turn on and off an IoT device, trigger a function, and determine whether a function is successfully executed.

\subsubsection{Router}

The router is configured using a standard NAT setup, with one network interface connected to the Internet and another one bridged to the IoT devices under test.
As part of the router's DHCP support, IoT devices are assigned a DNS server that we control (and that serves as a proxy for the ISP's DNS server). Together with traffic redirection rules and a \textit{dnsmasq} instance, our testbed intercepts all DNS requests, even if an IoT device uses a DNS server other than the DHCP-advertised one (\eg by using a public resolver).  
%
We collect all network traffic traversing the testbed using \textit{tcpdump}. 
The router can block IoT traffic destinations by IP address (including IP masks) and by altering DNS responses whose request matches a given pattern. 
When a DNS destination is blocked, it is resolved as localhost (127.0.0.1).

\subsubsection{Support Scripts}
\label{sub:scripts}

\new{We use support scripts to power on/off the devices, to \emph{trigger} their functions, and to \emph{probe} them to find out if a function execution is successful or not. The invocation of these scripts is fully automated, as part of our automated experiments methodology. 
Please note that while every step of every experiment (including the invocation of support scripts) is fully automated to allow our approach to scale, the creation of support scripts requires programming effort, which is a manual process that is device-dependent and functionality-dependent.  
However, once support scripts are written, they can be reused across experiments and only need to be rewritten after major changes in the device interaction interface.}

\noindent \textbf{Power on/off Scripts.}
The IoT devices are plugged into programmable smart plugs, and we use scripts to turn these smart plugs on and off so that we can reset the IoT devices by power-cycling them after every test.

\noindent \textbf{Trigger Scripts.}
To scale our analysis to many devices, we automate interactions with IoT devices by triggering their functions programmatically. We call the different automation strategies \emph{device triggers}, which are function and device-specific.
\new{An example of trigger for turning on a smart bulb is to programmatically provide input to its companion app in such a way to turn it on.}

\noindent \textbf{Probe Scripts.}
To verify that a trigger correctly executes a function, our methodology relies on some additional scripts, called \emph{device probes}, which programmatically query the status of a device, or analyze any signals it produces. The scripts then compare this information against ground truth, to check whether the execution of a function was successful or not.
\new{Probe scripts are also device-specific. An example of probe to determine if a light bulb is on is to retrieve the screenshot of its companion app and compare it to a previously retrieved screenshot where we know that the bulb was on.}

\subsection{Functionality Experiments}

 \begin{figure}
  \centering
  \includegraphics[width=0.8\linewidth]{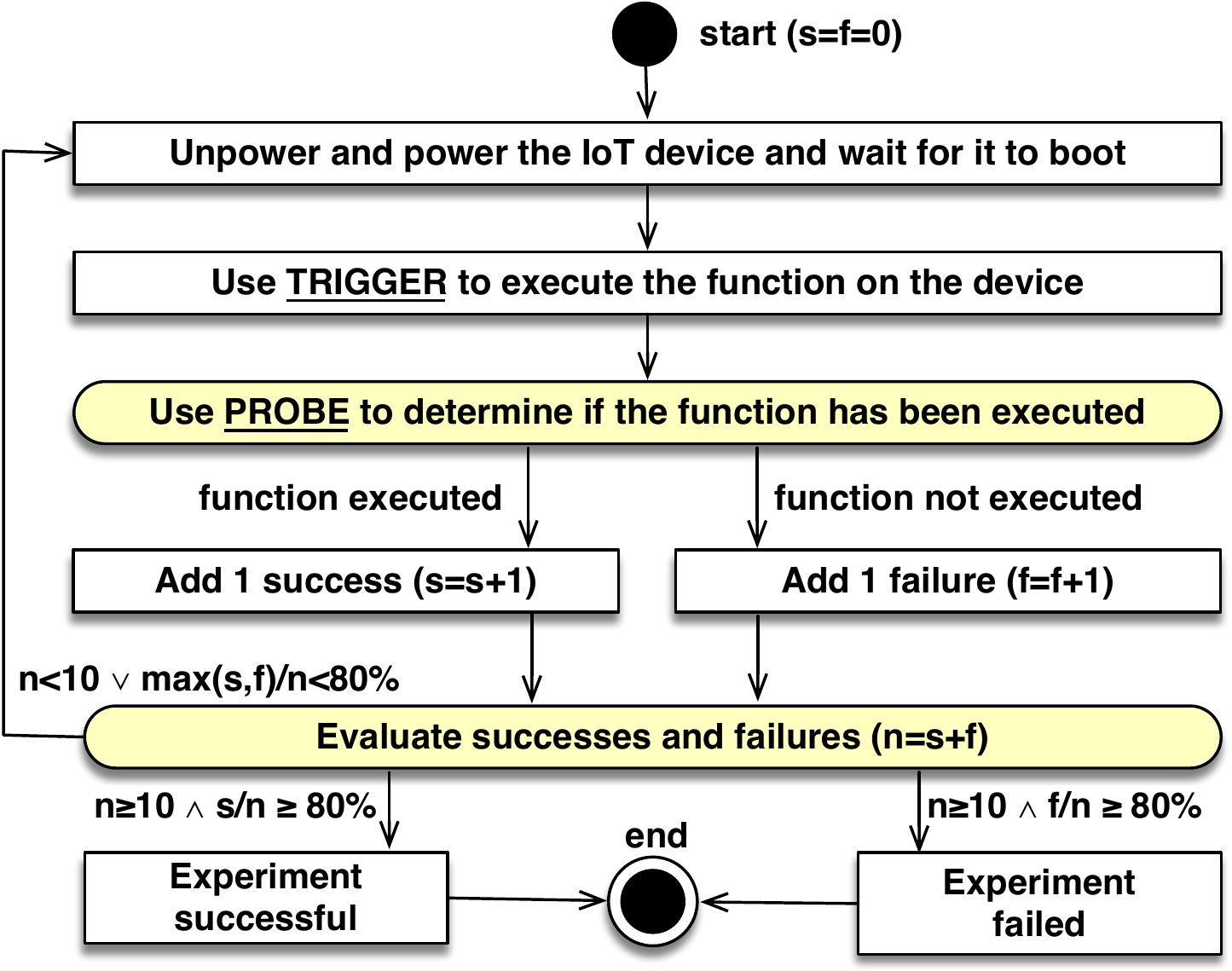}
  \caption{\new{Functionality experiment}. The algorithm iterates the execution of a function at least 10 times: $s$, $f$, $n$ are counters for successful, failed, and total iterations. When an 80\% consensus is achieved the algorithm terminates with a success or fail result. }
  \label{fig:experiment}
\end{figure}

The basic unit of our methodology is the \emph{functionality experiment} (see Fig.~\ref{fig:experiment}).
We define it as the \new{fully automated} process to verify if a function of an IoT device can be executed or not.
During our preliminary experiments we found that \new{several} IoT devices are less than 100\% reliable in terms of correctly executing their functionality in normal operating conditions (see our evaluation in \S\ref{sub:probesevaluation}).
The reasons are various, and span from random reboots, to the temporary disruptions in connectivity to cloud services. 
To cope with such events that prevent 100\% accurate probe scripts, we choose to use probe scripts as long as they have at least 80\% accuracy (\ie they can be incorrect at most 20\% of the time due to whatever reason, including the device not behaving correctly).
To ensure that we can reliably use information from probe scripts that may be inaccurate, we test the function \emph{multiple times} (at least 10), and we consider the result correct if a strong majority (at least 80\%) of the test results are consistent.
\new{This ensures that the whole functionality experiment has a negligible probability of an incorrect result:
while the odds of any one test failing can be significant, the odds that all of the multiple tests failing is substantially lower.}

Specifically, each functionality experiment iterates at least 10 times the following three steps. 

\noindent \textbf{Step 1.} We power on the device using the testbed's power script (to turn on the smart plug powering the device) and then wait for it to finish booting. The wait time is determined empirically using probe scripts, and we have found that a two-minute delay is enough for all the devices we tested. 

\noindent \textbf{Step 2.} We trigger the function of the experiment by invoking the proper device trigger.

\noindent \textbf{Step 3.} We use the device probe to verify that the function has been actually executed: if it is, we report the iteration as successful, otherwise as failure.
If, after all the iterations, at least 80\% of them is successful, we report the experiment as successful; if at least 80\% of the iterations fails, we report the experiment as a failure. 

If 80\% of the iterations is neither a success or a failure, we run an additional iteration and evaluate this test again.
The algorithm keeps performing iterations until the 80\% threshold is reached. 
If the threshold is never reached the algorithm would not terminate (\ie it keeps iterating with no success or fail result); however since we assume probes that are at least  80\% accurate (as it will be shown in \S\ref{sub:probesevaluation}), having a threshold of 80\% ensures that we achieve convergence in the long run.

Given that the probes have a maximum 20\% probability to produce an incorrect result during an iteration, there is a chance that a functionality experiment terminates with an incorrect result; however, such a chance is negligible (less than 0.0078\%\footnote{This upper bound for the probability of an incorrect result for our algorithm has been calculated by considering that the number of incorrect results of a probe (over $n$ iterations) follows a binomial distribution with parameters $n$ and $p=0.2$.}) since the incorrect result must happen in at least 80\% of the iterations.
\new{During} our experiments the algorithm always converges, meaning that the probes fulfilled their accuracy requirement. 

\subsection{Building the List of Destinations}
\label{sub:list}

To determine what destinations are required or not for a given IoT function, we need to first obtain the list of destinations during a preliminary destination-observing experiment, which consists of running a functionality experiment (which is composed of minimum 10 iterations of function invocations) without blocking any network traffic, and collecting the list of destinations contacted by the device.
All destination-observing experiments under normal circumstances are successful since we do not block any traffic.
We primarily identify all contacted destinations by hostname rather than by IP address. 
To do this, for each IP destination, we look at all the DNS traffic for the device to find the DNS hostname that resolved to the IP address.
If the hostname cannot be determined using this method, we simply use the IP address as the destination.
\new{We exclude from this process DNS (TCP/UDP port 53) and NTP (network time protocol, UDP port 123) destinations. We always allow these protocols since they are needed to resolve hostnames (DNS) and synchronize device clocks (NTP) for checking TLS certificate validity.}

\new{Many cloud services for IoT devices use replicated servers that provide the same functionality, and they sometimes use different (but similar) DNS names and IP addresses for each replica. To facilitate analysis and streamline blocklists, we \emph{group} destinations that are \emph{ephemeral}, \ie they appear in less than 80\% of the iterations of the destination-observing experiments.
For example, if \texttt{a.zz.com} is an ephemeral destination contacted in half of the iterations, and \texttt{b.zz.com} is contacted in the other half, they are both replaced by a single destination group \texttt{*.zz.com}, which appears in 100\% of the iterations.
All ephemeral destinations encountered in our experiments were successfully replaced with second-level wildcard domains.
For more details on this process, see Appendix~\ref{sec:grouping}.}

\subsection{Determining Required Destinations}

 \begin{figure}
  \centering
  \includegraphics[width=0.9\columnwidth]{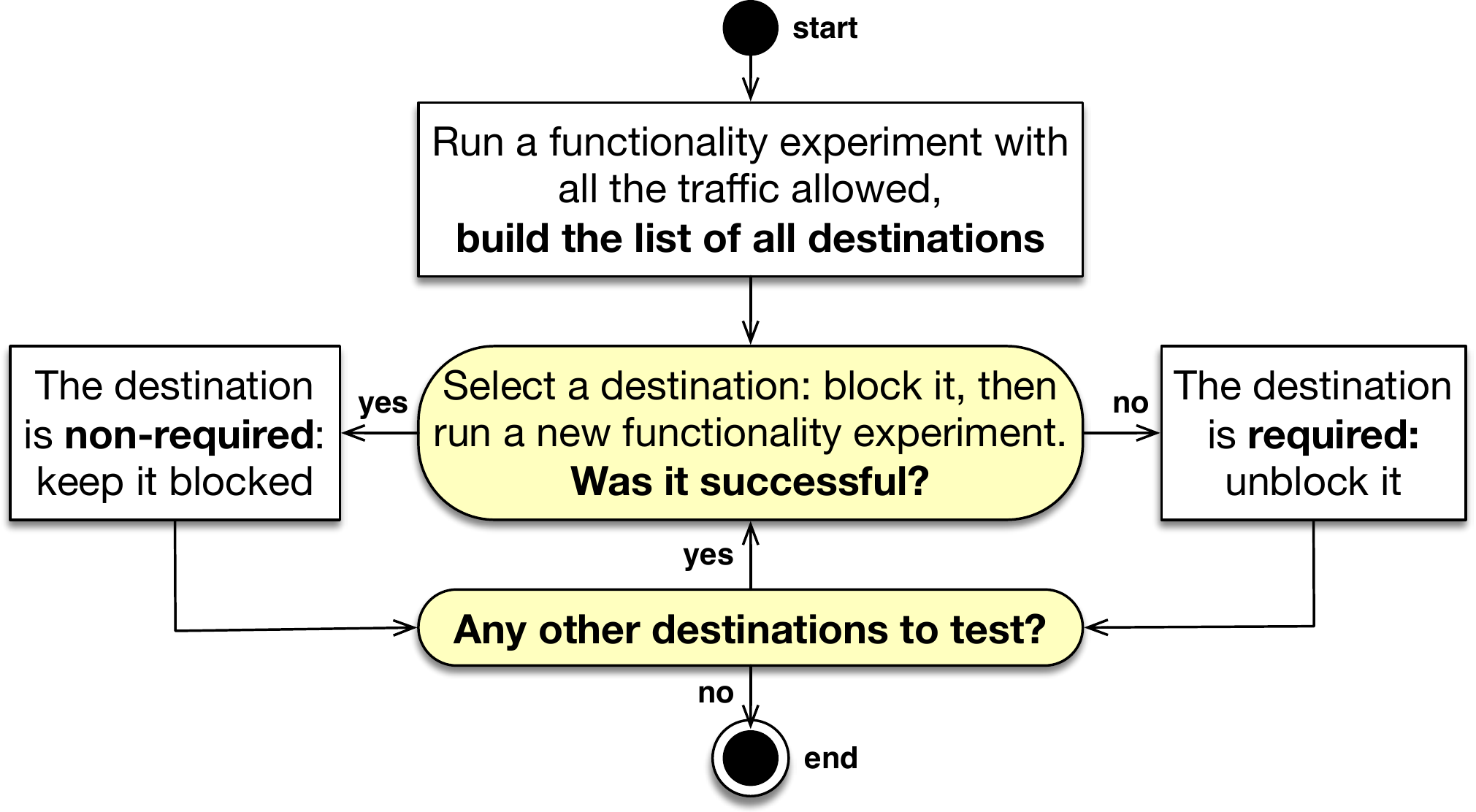}
  \caption{Methodology for detecting required destinations.}
  \label{fig:method}
\end{figure}

The algorithm for creating the list of required and non-required destinations is reported in Fig~\ref{fig:method}.

\noindent \textbf{Step 1}. \emph{Building the list of destinations.}
See~\S\ref{sub:list}. 

\noindent \textbf{Step 2}. \emph{Marking each contacted destination.}
Iteratively test all contacted destinations by running a functionality experiment for each of them.
In each iteration, the considered destination is blocked.
If the experiment succeeds, such destination is marked as non-required, and will stay blocked.
If the experiment fails, such destination is marked as required, and will be unblocked.
This process repeats until all destinations collected during the first step are classified.
\section{IoT Devices}
\label{sec:devices}

In this section we describe the IoT devices we use in our experiments, and all specialized device triggers and probes we use to apply the methodology described in~\S\ref{sec:method}.

\subsection{List of Devices and Tested Functions}
\label{sub:devices}

The devices we consider are consumer IoT devices typically deployed in a smart home.
We have chosen devices under these categories (see the first column of Table~\ref{table:probe}):
\begin{packed_itemize}
\item \emph{Camera}. Devices equipped with a camera sensor, such us smart camera systems and smart doorbells. The function we test is to watch a live stream.
\item \emph{Home automation and appliances}. Devices that offer home automation capabilities such as smart lights, kitchen appliances. The function we test is switching the device on and off.
\item \emph{Smart hubs}. Devices coordinating other non-IP IoT devices (\ie Zigbee). The function we test is switching the devices on and off.
\item \emph{Smart speakers}. Speakers that offer a voice assistant. We test responses to the voice command, ``What is the capital of Italy?''
\item \emph{Video.} Devices designed to stream video on a TV. We test streaming from YouTube.
\end{packed_itemize}

\noindent The criteria we use for choosing the function to test are:
(\emph{i}) it must be a function that is characteristic of the device category;
(\emph{ii}) it must be intended for user-initiated interactions and not initiated by the device itself;
(\emph{iii}) it must be amenable to triggers and probes.

To better represent how IoT devices behave in the wild, we try to keep their default configuration and privacy settings unaltered and we do not perform user-initiated firmware \new{updates}.
Devices are still allowed to perform automated firmware \new{updates} when such a feature is enabled in the default configuration.


\begin{table}[t]
	\centering
		\resizebox{1.0\columnwidth}{!}{	
		\begin{tabular}{lrccccc}
			\multirow{3}{*}{\textbf{Device}} &\multirow{1}{*}{\textbf{Trigger}} &\multirow{1}{*}{\textbf{Probe}} & \multirow{1}{*}{\textbf{Success}}& \multirow{1}{*}{\textbf{Success}} & \multirow{1}{*}{\textbf{Success}} \\
			 & & & (May)& (July) & (October) \\
			
\hline
\hline
Camera (Watching live)& &&\\	
\hspace{3mm} Blink & App  & Screen &  93.3\% & 96.7\% & 100\% \\
\hline
\hspace{3mm} Bosiwo & App  & Screen &  100\% &100\% & 100\% \\
\hline
\hspace{3mm} iCSee & App  & \sout{Screen} &  80\% &76.7\% & 70\%  \\
\hline
\hspace{3mm} Reolink & App  & \sout{Screen} &  70\% & 70\% & 60\%  \\
\hline
\hspace{3mm} Wansview & App  & Screen &  90\% &93.3\% & 100\%  \\
\hline
\hspace{3mm} Yi & App  & Screen &  100\% &100\% & 100\%  \\
\hline
\hline
Home-automation (Switching on/off) &&&\\
\hspace{3mm} App Kettle & App  & Screen &  100\% &100\% & 100\%  \\
\hline
\hspace{3mm} Honeywell thermostat & App  & Screen &  100\% &100\% & 100\%   \\
\hline
\hspace{3mm} Magichome & App  & Screen &  100\% &100\% & 100\%  \\
\hline
\hspace{3mm} Meross dooropener & App  & Screen &  100\% &100\% & 100\%  \\
\hline
\hspace{3mm} Nest thermostat & App  & Screen &  100\% &100\% & 100\%   \\
\hline
\hspace{3mm} Netatmo weather & App  & Screen &  100\% &100\% & 100\%  \\
\hline
\hspace{3mm} Smarter coffee machine& App  & Screen &  100\% &100\% & 100\% \\
\hline
\hspace{3mm} Smartlife bulb & App  & Screen &  100\% &100\% & 100\%  \\
\hline
\hspace{3mm} Smartlife remote control & App  & Screen &  100\% &100\% & 100\%  \\
\hline
\hspace{3mm} Sousvide cooker & App  & Screen &  100\% &100\% & 100\%  \\
\hline
\hspace{3mm} Switchbot & App  & Screen &  100\% &100\% & 100\% \\
\hline
\hspace{3mm} TP-Link bulb & App  & Screen &  100\% &100\% & 100\% \\
\hline
\hspace{3mm} TP-Link plug & App  & Screen &  100\% &100\% & 100\%  \\
\hline
\hspace{3mm} Wemo plug & App  & Screen &  100\% &100\% & 100\%     \\
\hline
\hspace{3mm} Xiaomi rice-cooker & App  & \sout{Screen} &  46.7\% &53.3\% & 40\% \\
\hline
\hline
Smart-hub (Switching on/off)& &&\\
\hspace{3mm} Insteon & App  & Screen &  100\% &100\% & 100\%  \\
\hline
\hspace{3mm} Lightify & App  & Screen &  100\% &100\% & 100\%  \\
\hline
\hspace{3mm} Philips & App  & Screen &  100\% &100\% & 100\%    \\
\hline
\hspace{3mm} Samsung & App  & Screen &  96.7\% &100\% & 100\%   \\
\hline
\hspace{3mm} Sengled & App  & Screen &  100\% &100\% & 100\% \\
\hline
\hline
Smart-Speaker (Asking questions)&&&\\
\hspace{3mm} Allure & Voice  & Traffic &  100\% &100\% & 100\%   \\
\hline
\hspace{3mm} Echo Dot & Voice  & Traffic &  100\% &100\% & 100\% \\
\hline
\hspace{3mm} Google Home & Voice  & Traffic &  100\% &100\% & 100\% \\
\hline
\hline
Video (Watching YouTube)&&&\\
\hspace{3mm} Fire TV & App  & Traffic &  100\% &100\% & 100\% \\
\hline
\hspace{3mm} Roku TV & App  & Traffic &  100\% &100\% & 100\%  \\
\hline
\end{tabular}
}
\vspace{-0.5mm}
\caption{List of our devices by category. For each of them: triggering and probing strategy we used, and probe success rate evaluation in three different point in time (May, July, and October 2020). \new{Crossed out probing strategies are the ones we could not use programmatically due to insufficient success rate (see \S\ref{sub:probesevaluation}).}} 
\label{table:probe}
\end{table}

\subsection{Specialized Device Triggers}
\label{sub:triggers}

As discussed in~\S\ref{sub:scripts}, we use device-dependent trigger scripts to execute functionality.
The triggering strategies we use for each device are reported in the second column of Table~\ref{table:probe} and described as follows.

\noindent \textbf{Companion app.}
This triggering strategy is possible for IoT devices that can be controlled via a companion app compatible with Android.
We install this app on an Android phone that is \emph{not on the same LAN} as the IoT device (to force the communication to happen over the Internet rather than directly), and then trigger each function by emulating user interactions programmatically using the Android Debug Bridge.

\noindent \textbf{Voice assistant.}
This strategy is used for smart speakers.
We use the Google voice synthesizer connected to a set of regular speakers (placed next to the smart speaker) to programmatically issue voice commands.

\begin{figure}
  \centering
  \includegraphics[width=0.85\linewidth]{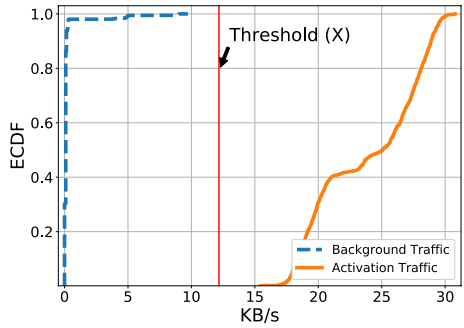}
  \caption{ECDF of the data peaks $p_{i,j}$ over 135 experiments for the Echo Dot: the plot shows a clear distinction between the data peaks when its main function is executed (activation data peaks) and when it is not (background data peaks).}
  \label{fig:trafficpatterns}
\end{figure}

\subsection{Specialized Device Probes}
\label{sub:probes}

We use several device-dependent strategies also for probing the devices;
the probing strategies we use for each device are reported in the third column of Table~\ref{table:probe} and described as follows.

\noindent \textbf{Companion app screenshot.}
For the majority of IoT devices, a companion app is available as a method to obtain their state.
For this method, we use a preliminary experiment during which we take sample screenshots of the app for each device, and we use this as the ground truth for the correct state after executing a function. For subsequent experiments, we take screen shots after using a trigger, and compute the similarity of each screenshot to the ground truth to infer the state of the device. 

To quantify this similarity, we simply check how many pixels differ in the two screenshots by more than a particular threshold. We use the same parameters for all companion apps, \new{which we tuned} through an analysis on a few sample screenshots.

\noindent \textbf{Network traffic patterns.}
This probe analyzes the patterns of the network traffic generated by an IoT device in two situations: when the function has been executed (\emph{activation traffic}) and when the function was not executed (\emph{background traffic}).
During preliminary experiments we observed that when the main function is executed for some devices (typically streaming devices such as smart speakers), they significantly increase the amount of data transmitted to certain destinations compared to when the function is not executed (see Echo Dot example in Fig.~\ref{fig:trafficpatterns}).
Based on this observation, we automatically detect traffic bursts corresponding to the traffic pattern for the main function of a device. 

Specifically, for device $i$ and an experiment $j$, we consider the data peak $p_{i,j}$, defined as the maximum amount of traffic sent by the device to such destinations among all 20-second window samples over the full duration of the experiment.
From a series of preliminary experiments where we know as ground truth that the main function of device $i$ is executed and not executed, we calculate the constants $A^{\textit{min}}_i$ and $B^{\textit{max}}_i$, where $A^{\textit{min}}_i$ (minimum activation peak) is the minimum data peak $p_{i,j}$ over all experiments $j$ with execution, and $B^{\textit{max}}_i$ (maximum background peak) is the maximum data peak $d_{i,j}$ over all experiments $j$ \emph{without} execution.
We then define the data peak activation threshold $X_i$, as the average between $A^{\textit{min}}_i$ and $B^{\textit{max}}_i$: any data peak that is larger than this threshold signals the presence of activation traffic.

The probe then uses $X_i$ to determine whether device $i$ had its function executed or not during a new experiment $k$:
if $p_{i,k}>X_i$ (\ie the experiment has a data peak that is larger than the peak activation threshold), the probe returns success for $k$, otherwise it returns failure.
Table~\ref{tab:threshold} shows the destinations and parameters for the network traffic probes, calculated over a minimum of 135 preliminary experiments for each device.


\begin{table}[t]
\begin{center}
\resizebox{0.8\columnwidth}{!}{
 \begin{tabular}{c c c c c} 
 Device ($i$) & Destination & $B^{max}_i$ & $A^{min}_i$&Threshold ($X_i$) \\
  &   & [KB/s] &[KB/s] &[KB/s] \\
 \hline\hline
Allure  &  bob-dispatch-*& 6.763 & 13.342 & 10.052 \bf{}\\
			& *.amazon.com  &  & &  \\ 
 \hline
 Echo Dot &  bob-dispatch-*& 8.889 & 15.455 &  \bf{} 12.172\\
 (\third{} gen.) & *.amazon.com  &  & &  \\ 
\hline
Fire TV &  youtube.com  & 0 &109.38 &  \bf{}54.69\\
 \hline
Google Home &  google.com  & 41.69 &50.483 &  \bf{}46.086\\
 \hline
 Roku TV &  youtube.com  & 0 &140.364 &  \bf{}70.182\\
 \hline
\end{tabular}
}
\end{center}
\vspace{-0.5mm}
	\caption{Network traffic probe thresholds for data peaks. $B^{max}_i$ is the maximum peak in background traffic (\ie no functionality execution), $A^{min}_i$ is the minimum peak in activation traffic (\ie with functionality execution),  $X_i$ is the data peak threshold, \ie the minimum peak required for detecting device activation.}	\label{tab:threshold}
\end{table}


\subsection{Probes Evaluation}
\label{sub:probesevaluation}

\noindent \textbf{Probes Evaluation Method.}
Our method for classifying required destinations relies on probes that are at least 80\% accurate on average.
To identify whether this property holds, we run 70 \emph{probe evaluation experiments} per device in three points in time (10 in May, 30 in July, and 30 in October 2020). 
Each probe evaluation experiment is a set of functionality experiments run in the following three situations: (\emph{i}) with all the destinations allowed, where we know \emph{a priori} that the function execution succeeds (\ie testing to see if the probe correctly detects successful experiments); (\emph{ii}) with all the destinations blocked, where we know \emph{a priori} that the function fails (to ensure that the probe detects experiment failures); (\emph{iii}) with all the destinations allowed, but without executing the trigger, to test whether the probe detects that the function \new{is} not executed.

Once the experiments are complete, we calculate the success (failure) rate of the probe, defined as the number of correct (incorrect) probe results over the total number of experiments for that probe.
We consider the minimum between the success rate and the failure rate as a conservative metric to measure the accuracy of a probe, and use it as the expected probability to provide a correct result.

\noindent \textbf{Detecting successes.}
The results of our probes evaluation method for detecting the success of a function execution are reported in the last three columns of Table~\ref{table:probe},
where each column represents the evaluation at a different point in time.
For 28 of \numdev{} devices, our probes correctly and consistently recognize the execution of a function
in at least 80\% of the cases, which satisfies the requirement of our method for classifying destinations.
For the remaining three devices (iCSee Doorbell, Reolink Camera, and Xiaomi Rice Cooker), we could not find probes that \new{are} at least 80\% accurate during all the three points in times. 

\noindent \textbf{Detecting failures.}
We find that our probes correctly recognize a function execution failure both for cases where all traffic is blocked, and when the function is (intentionally) not triggered. As a result, it is very unlikely that a probe will report as successful an experiment where the execution of a function fails, since this kind of error never happened during our 4,340 probe evaluation experiments. 

\noindent \textbf{\new{Dealing with inaccurate probes.}}
\new{For the three devices whose probes are not accurate enough, we cannot use our fully automated analysis approach because  we do not have an automated way to detect if a trigger is successful. 
To still include them in our study, we probe their status manually, while keeping all the remaining steps of our approach automated. }

\section{Identifying Non-essential Traffic}
\label{sec:destinations}

\begin{figure}[t]
	\begin{center}
		\includegraphics[width=0.7\columnwidth]{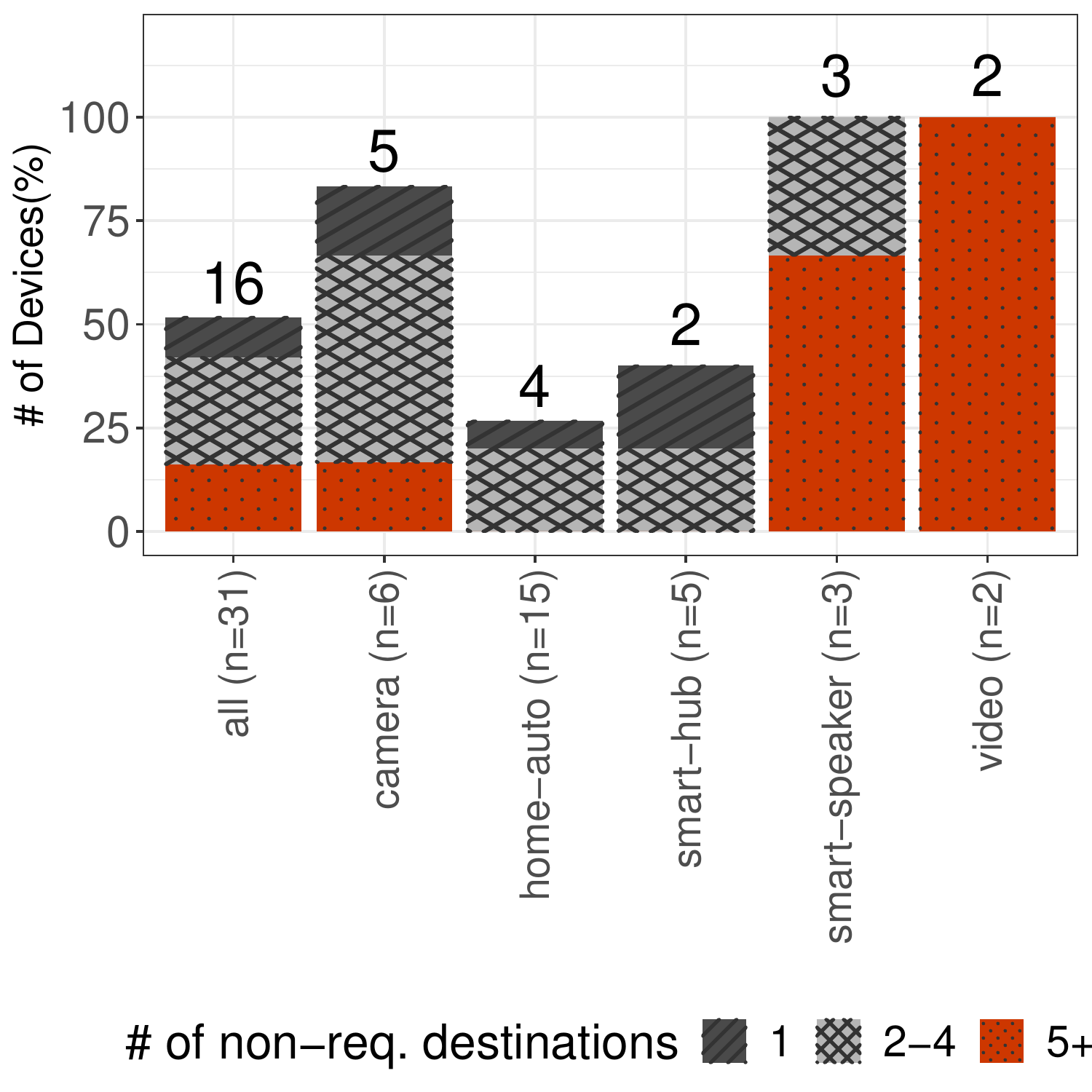}
		\caption{Percentage of devices with at least one non-required destination. Sub-bars show how the number of non-required destinations is distributed among the devices for each category. }
		\label{fig:percnon-req}
	\end{center}
\end{figure}


\begin{table*}[h!]
	\resizebox{1\textwidth}{!}{
		\centering
		\begin{tabular}{m{0.01\textwidth}p{0.15\textwidth}||c|c|c|p{0.8\textwidth}}
		\centering
			&\textbf{Device} & \textbf{Dest. \#} & \textbf{Req. \#}& \textbf{Non-Req. \#} & \textbf{List of Non-Required Destinations}\\ 
			\hline
			\hline
\multirow{5}{*}{\rotatebox[origin=c]{90}{\small Camera}} &\hspace{1mm} Bosiwo & \centering 4 & \centering2 &\centering2& \color{gray}\textit{54.157.82.107}, \color{red}\underline{210.72.145.44}\\
\cline{2-6}
& \hspace{1mm} iCSee & \centering6   & \centering 2 &\centering4& \color{gray}\textit{47.52.222.172, 47.52.32.118, api.gdxp.com, oss-us-west-1.aliyuncs.com} \\
\cline{2-6}
&\hspace{1mm} Reolink & \centering2    & \centering 1 &\centering1 & pushx.reolink.com   \\
\cline{2-6}
&\hspace{1mm} Wansview & \centering 9  & \centering 3 &\centering6&  \color{gray}\textit{159.65.95.225, 3.122.229.130 , ajcloud.net, htpdate.ajcloud.net, sdc-isc.ajcloud.net, *.backblaze.com}  \\
\cline{2-6}
&\hspace{1mm} Yi & \centering5   &  \centering3 &\centering2  &api.eu.xiaoyi.com, log.eu.xiaoyi.com  \\
\hline
\hline
\multirow{3}{*}{\rotatebox[origin=c]{90}{\small Home-auto}} & \hspace{1mm} Nest thermostat & \centering3&  \centering 2 &\centering 1& frontdoor.nest.com  \\
\cline{2-6}
& \hspace{1mm} TP-Link bulb & \centering4 &  \centering1 &\centering3&  euw1-api.tplinkra.com, n-deventry.tplinkcloud.com, use1-api.tplinkra.com \\
\cline{2-6}
& \hspace{1mm} TP-Link plug &\centering 4 & \centering 1  & \centering3& euw1-api.tplinkra.com, n-deventry.tplinkcloud.com, use1-api.tplinkra.com \\
\cline{2-6}
& \hspace{1mm} Xiaomi rice-cooker& \centering7   &  \centering3 &\centering4& \color{gray}\textit{183.84.5.203, 58.83.160.36, 123.125.102.215, 110.43.0.83}   \\
\hline
\hline
\multirow{2}{*}{\rotatebox[origin=c]{90}{\small Hub}} & \hspace{1mm} Philips & \centering4   &  \centering2 &\centering2&  diagnostics.meethue.com, ecdinterface.philips.com\\
\cline{2-6}
& \hspace{1mm} Samsung & \centering3    & \centering 2 &\centering1 & fw-update2.smartthings.com \\
\hline
\hline
\multirow{4}{*}{\rotatebox[origin=c]{90}{\small Speaker}} & \hspace{1mm} Allure& \centering3  &   \centering 1 &\centering2&  \color{gray}\textit{api.amazon.com, d1enchupjctwud.cloudfront.net} \\
\cline{2-6}
& \hspace{1mm} Echo Dot & \centering10  &   \centering 3 &\centering7& arcus-uswest.amazon.com, *.cloudfront.net, device-metrics-us.amazon.com, dp-gw.amazon.com, fireoscaptiveportal.com, \color{gray}\textit{prod.amcs-tachyon.com}\color{black}, s3-1-w.amazonaws.com\\
\cline{2-6}
& \hspace{1mm} Google Home & \centering9    &\centering  4 &\centering5&   youtube-ui.l.google.com, clientservices.googleapis.com, fcm.googleapis.com, *.googlevideo.com, storage.googleapis.com\\
\hline
\hline
\multirow{6}{*}{\rotatebox[origin=c]{90}{\small Video}} & \hspace{1mm} Fire TV & \centering14  & \centering  3 & \centering11&\color{red}\underline{aax-eu.amazon-adsystem.com}, \color{black}arcus-uswest.amazon.com, bob-dispatch-prod-eu.amazon.com, *.cloudfront.net, device-metrics-us.amazon.com, api.amazon.com, ktpx-eu.amazon.com, \color{red}\underline{api-global.eu-west-1.prodaa.netflix.com}\color{black}, mas-ext-eu.amazon.com, mas-sdk.amazon.com, msh.amazon.com\\
\cline{2-6}
& \hspace{1mm} Roku TV& \centering10    &\centering  2 &\centering8&    \color{red}\underline{api-global.eu-west-1.prodaa.netflix.com}\color{black}, configsvc.cs.roku.com, cooper.logs.roku.com, \color{red}\underline{customerevents.eu-west-1.prodaa.netflix.com}, \color{red}\underline{ichnaea.eu-west-1.prodaa.netflix.com}, \color{red}\underline{partnerad.l.doubleclick.net}\color{black}, scribe.logs.roku.com, \color{red}\underline{uiboot.eu-west-1.prodaa.netflix.com}\\
\hline
\hline
& Other devices (15) &\centering 22 &\centering22&\centering 0 &   \\ 
\hline
\hline
\textit{Total} &\centering  \textit{31} &\centering  \textit{119} &\centering \textit{57} &\centering \textit{62} &    \\ 
		\end{tabular}
	}
	\vspace{-0.5mm}
	\caption{Non-required destinations. We report, for each device (having at least one non-required destination), the total number of destinations, the number of required destinations, the number of non-required destinations, and the list of non-required destinations. Colors identify the destination party type (see~\S\ref{sub:deviceparty}): first party, \color{gray}\textit{support party}\color{black}, and \color{red}\underline{third party}\color{black}. For a version of this table, which includes all our IoT devices and the list of required destinations as well, see Table~\ref{tab:reqnon-reqdestinations} in Appendix~\ref{sec:appendix}. }
	\label{tab:non-reqdestinations}
\end{table*}

\begin{figure*}[t]
    \centering
    \begin{minipage}{0.44\textwidth}
        \centering
        \includegraphics[width=1\linewidth]{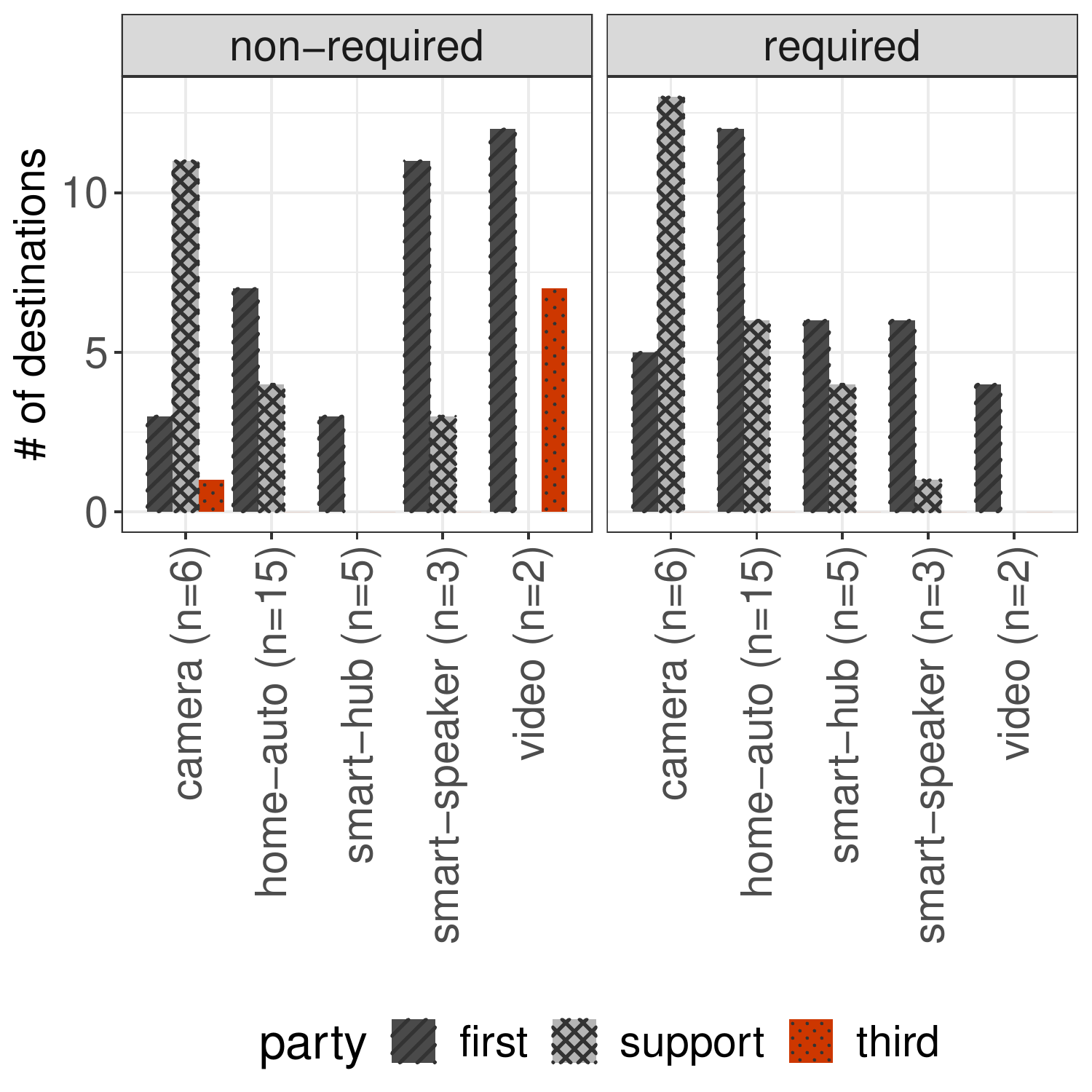}
    \end{minipage}%
    \hspace{+50pt}
    \begin{minipage}{0.44\textwidth}
        \centering
        \includegraphics[width=1\linewidth]{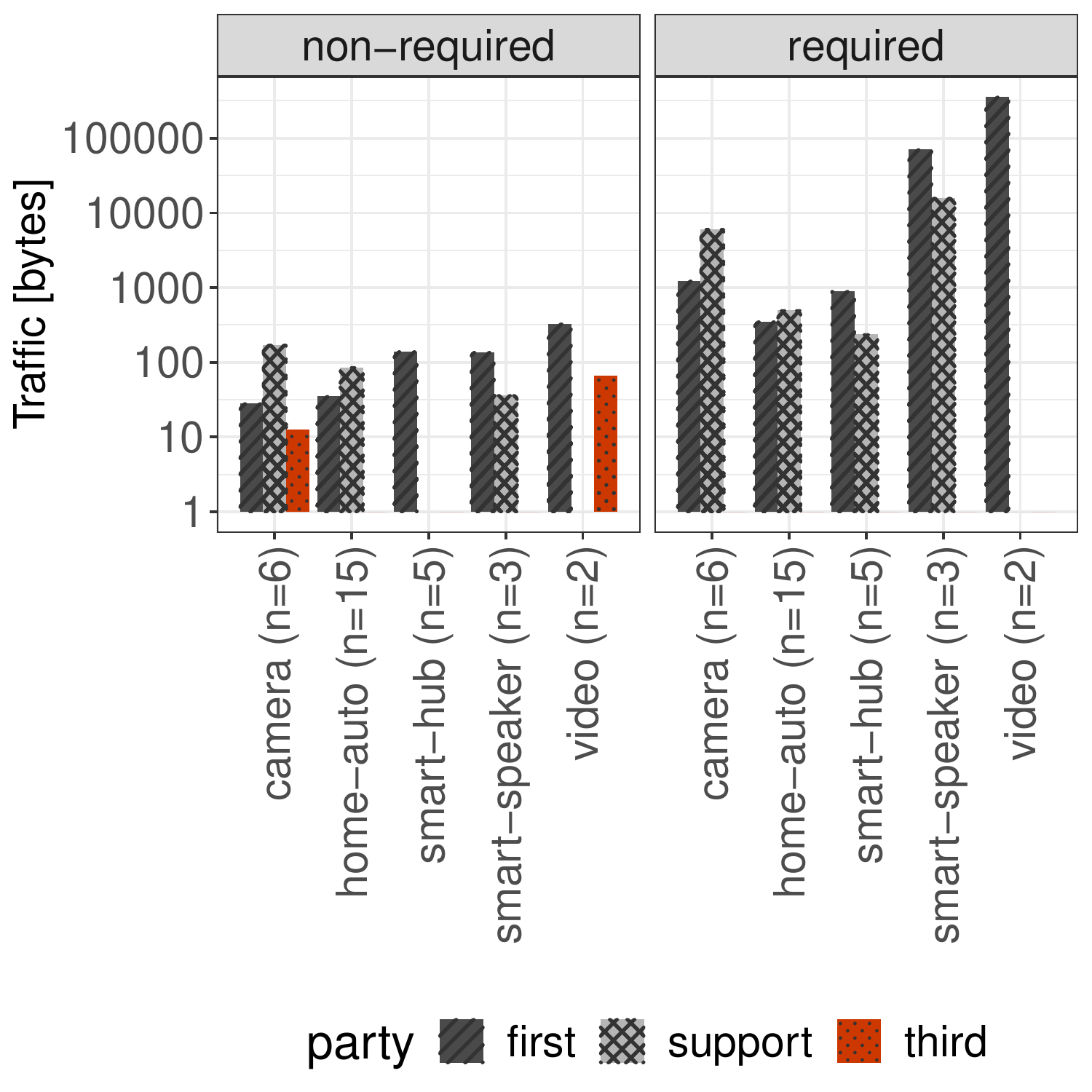}
    \end{minipage}
     \caption{Number (left) and total traffic (right) of required/non-required destinations. The total traffic is the average payload (without headers) produced by the devices during successful function invocations.}
               \label{fig:bar_dest_party_cat}
\end{figure*}

We answer our second research question by applying the methodology described in \S\ref{sec:method} to identify and characterize non-essential traffic produced by our IoT devices (\S\ref{sec:devices}).

\subsection{Impact of Device Category}
\label{sub:category}

We now characterize the destinations of the traffic of each device in terms of traffic sent to required destinations (essential traffic) and traffic sent to non-required destinations (non-essential traffic).
Fig.~\ref{fig:percnon-req} shows that 52\% of the IoT devices we tested produce non-essential traffic, with 25\% and 16\% of them contacting respectively at least 2 and 5 non-required destinations. 
The figure shows that all devices tested in the smart speakers and video categories produce non-essential traffic.

We now consider these devices and non-required destinations in more detail (Table~\ref{tab:non-reqdestinations}). Of these devices, the Amazon Fire TV (11 destinations) and the Roku TV (8 destinations) contact the largest number of non-required destinations.
Notably, devices in the camera category also contact up to six non-required destinations---a surprising result since such devices do not have third-party apps or UIs that can include non-required traffic such as advertising. 
In most of these notable cases the number of non-required destinations tend to be larger than the number of required ones.
Finally, we observe several non-required destinations also on simple devices, such as the TP-Link plug, which sends non-essential traffic to three non-required destinations while only having a single required destination.

On the other hand, \numdevnoblockable{} of \numdev{} devices contact required destinations only, \numreqdestnoblock{} in total. Table~\ref{tab:reqnon-reqdestinations} in  Appendix~\ref{sec:appendix} lists all the required and non-required destinations.

\noindent \textbf{Takeaways.}
Our results show that traffic to non-required destinations is present across all categories of IoT devices, and these devices often contact more non-required destinations than required ones. We further find that devices with a richer set of functions---such as smart speakers and video devices---are more likely to have such non-required traffic, followed by smart cameras.
\new{For the case of video devices, some of the non-required destinations are advertisers while others are related to video recommendations for pre-installed apps, a topic we discuss in \S\ref{sub:limitation}. Note that these destinations are not from background app activity, since it is disallowed for the 
Roku TV~\cite{RokuBackground, varmarken2020tv}, and we disabled background app activity on the Fire TV~\cite{firetvdev}.}
Regarding the cameras and other simpler devices, it is unclear why they produce non-essential traffic without the internal details of the devices and their software.

\subsection{Impact of Destination Party Type}
\label{sub:deviceparty}

In this section, we determine trends relating to whether a destination's party type (first party, support, third) are indicative of whether 
the destination is required or not for device functionality. \new{We use the same party type definitions and classification approach proposed in~\cite{moniotr} and we consider any advertising domain as third-party:  a \emph{First party} is a destination related to the device manufacturer or a related company responsible for fulfilling device functionality; a \emph{Support party} is any company providing outsourced computing resources such as CDN and cloud providers, which is not also a first party; a \emph{Third party} is a destination that is not a First party or a Support party. Third-party companies include advertising and analytics companies.} Table~\ref{tab:non-reqdestinations} uses text decoration in the rightmost column to indicate the party type for each observed device-destination pair.


To show aggregate findings, we group results by category in Fig.~\ref{fig:bar_dest_party_cat}, with the left figure analyzing the number of destinations and the right figure analyzing traffic volumes. We begin with the left plot, which plots how many required and non-required destinations are contacted for each destination party type and device category.
We find that third-party destinations are \emph{never} required, meaning that all their traffic is non-essential, while first and support parties are sometime\new{s} non-required and sometimes required. Overall, there are slightly more non-required first- and support-party destinations among the majority of devices and categories.

The right plot in Fig.~\ref{fig:bar_dest_party_cat} shows the average amount of data (payload only, no headers) that is sent to required and non-required destinations during successful function invocations. We observe that the vast majority of the data is sent to required destinations (\ie essential traffic) that are either first or support parties, while the volume \new{of} non-essential traffic is relatively small (\trafficblocked{} in total), and a mix of all the three party types.

For the devices we tested, non-essential traffic sent to third parties only occurs for the camera and video category, while all the non-essential traffic produced by IoT devices in the \new{smart-hub} category is sent to the first party.
In more detail, the third-party non-essential traffic is sent to advertisers such as \texttt{doubleclick.net} and \texttt{adsystem.com}, and other services such as  \texttt{netflix.com}, which are contacted by video devices, even if we do not use the Netflix app during our experiments. 

Most non-required destinations are first parties, with domain names suggesting that they are mainly used for logs, diagnostics and device configuration (\ie{}  \texttt{diagnostics.meethue.com},  \texttt{device-metrics-us.amazon.com}, \texttt{logs.roku.com}; see Table~\ref{tab:non-reqdestinations} for more detail).

\noindent \textbf{Takeaways.}
A key finding is that---for the devices we tested---third-party destinations are always non-required. 
This suggests that a simple blocking approach for such devices is simply to block all third-party communication.
\new{We also found that the video category has the largest number of third parties, some of which is explained by the menu screen loading previews of content from third-party apps.}

The fact that some non-required destinations are first or support party suggests that the manufacturer includes device activity that is unrelated to the main function. This could occur for good reasons such as firmware \new{updates}, or for more concerning reasons such as collection of device/user data.
Fortunately, only a small amount of payload is sent to non-required destinations, suggesting that the device is not exposing much information over these connections. 
On the other hand, we observe a significant number of non-required destinations contacted that are not first parties.
This is concerning because recent work shows that in such a small payload it is still possible to signal
the device presence, its status, and basic data from its
sensors~\cite{Trimananda2020, saidi2020haystack, Peek-a-Boo}, thus constituting a privacy and potential security risk. 

\subsection{Device-dependent Non-required Destinations}
\label{sub:devicedependent}


\begin{table}[t]
\begin{center}
\resizebox{1.0\columnwidth}{!}{
 \begin{tabular}{c c c} 
 Destination & Device (Non-required) & Device (Required)\\
 \hline\hline
api.amazon.com &  Allure &Echo Dot, FireTV\\
 \hline
bob-dispatch-prod-eu.amazon.com &  Fire TV & Echo Dot, Allure \\
\hline						
\end{tabular}
}
\end{center}
\vspace{-1mm}
	\caption{Device-dependant destinations. Destinations that are both required and non-required for different devices.}	
	\label{tab:dev-dep}
\end{table}

In this analysis we check whether any destinations that are non-required for a device are required for another device.
We define these destinations as \emph{device-dependent}. 
Knowing if there are any destinations that are device-dependent (under our definition) is important since their existence means that a blocking approach to prevent non-essential traffic cannot rely only on a flat list of destinations; rather blocking of destinations must be device-specific (requiring accurate device detection) for at least some devices.

Table~\ref{tab:dev-dep} shows the list of device-dependent destinations (first column), with the list of devices for which they are non-required (second column) and required (third column). We find two destinations that are non-required for some devices, but required for others, even for devices from the same manufacturer.
The first case is \texttt{api.amazon.com}, which is mandatory for Amazon devices to function, but not required by the Allure speaker, although it is powered by the same voice assistant of Amazon (Alexa).
The second case is \texttt{bob-dispatch.prod-eu.amazon.com}, which is required by all Amazon-powered smart speakers to process voice commands, but not required to watch YouTube on an Amazon Fire TV.


\noindent \textbf{Takeaways.}
On one hand, device-dependent destinations do exist, motivating blocklists that associate destinations to the actual device. 
However,
the function tested is also relevant in determining if a destination is required or not.
For example, Amazon Fire TV is primarily designed to stream TV through apps, but it also offers voice assistant functionality: for this reason the presence of a non-required destination typically used by Amazon-enabled smart speakers is not surprising.

\subsection{Common Non-required Destinations}
\label{sub:common}


\begin{table}[t]
\begin{center}
\resizebox{1.0\columnwidth}{!}{
 \begin{tabular}{c c c c c} 
 Destination & Device (Category) \\
 \hline\hline
*.cloudfront.net &  Echo Dot (Smart-speaker), FireTV (Video) \\
 \hline
api-global.eu-west-1.prodaa.netflix.com &  Fire TV (Video), Roku TV (Video) \\
  \hline
 arcus-uswest.amazon.com &  Echo Dot (Smart-speaker), Fire TV (Video) \\
\hline
 device-metrics-us.amazon.com &  Echo Dot (Smart-speaker), Fire TV (Video) \\
\hline
euw1-api.tplinkra.com &  TP-Link bulb, plug (Home-automation) \\
\hline
n-deventry.tplinkcloud.com &  TP-Link bulb, plug (Home-automation) \\
\hline	
use1-api.tplinkra.com &  TP-Link bulb, plug (Home-automation) \\
\hline						
\end{tabular}
}
\end{center}
\vspace{-0.5mm}
	\caption{Non-required destinations contacted by multiple devices.}	
	\label{tab:popular}
\end{table}

We now analyze non-required destinations that are in \emph{common} for the devices we tested, \ie non-required destinations contacted by more than one device.
The reason for this analysis is that, if multiple devices have the same non-required destination, such destination may have the same non-required purpose for other devices as well, which can help generalize our blocking approach. 

Table~\ref{tab:popular} reports the list of common non-required destinations (first column), and the devices/categories contacting them (second column).
We observe that the same manufacturers (\eg Amazon and TP-Link) have an overlap for non-required destinations. 
In the case of TP-Link, the non-required destinations contacted by a bulb and a plug coincide. 
Note that there is overlap from devices from different manufacturers in the video category: both Fire TV and Roku TV contact the same third-party service that is non-required (\texttt{api-global.eu-west-1.prodaa.netflix.com}).

\noindent \textbf{Takeaways.}
Our experiments demonstrate that devices from the same vendors tend to behave similarly, probably due to sharing some code among them and integrating them in the same IoT ecosystem.
This enables the extension of a blocking approach based on our destination lists to other or future devices from the same manufacturer.
Our analysis does not show notable situations in which devices from different vendors contact the same non-required destinations, except for the case of devices in the video category\new{, where both devices show Netflix video recommendations in their menu screen. }

\subsection{\new{Impact of SLDs, Protocols, and Ports}}
\label{sub:sld}

\begin{figure}[t]
	\begin{center}
		\includegraphics[width=0.94\columnwidth]{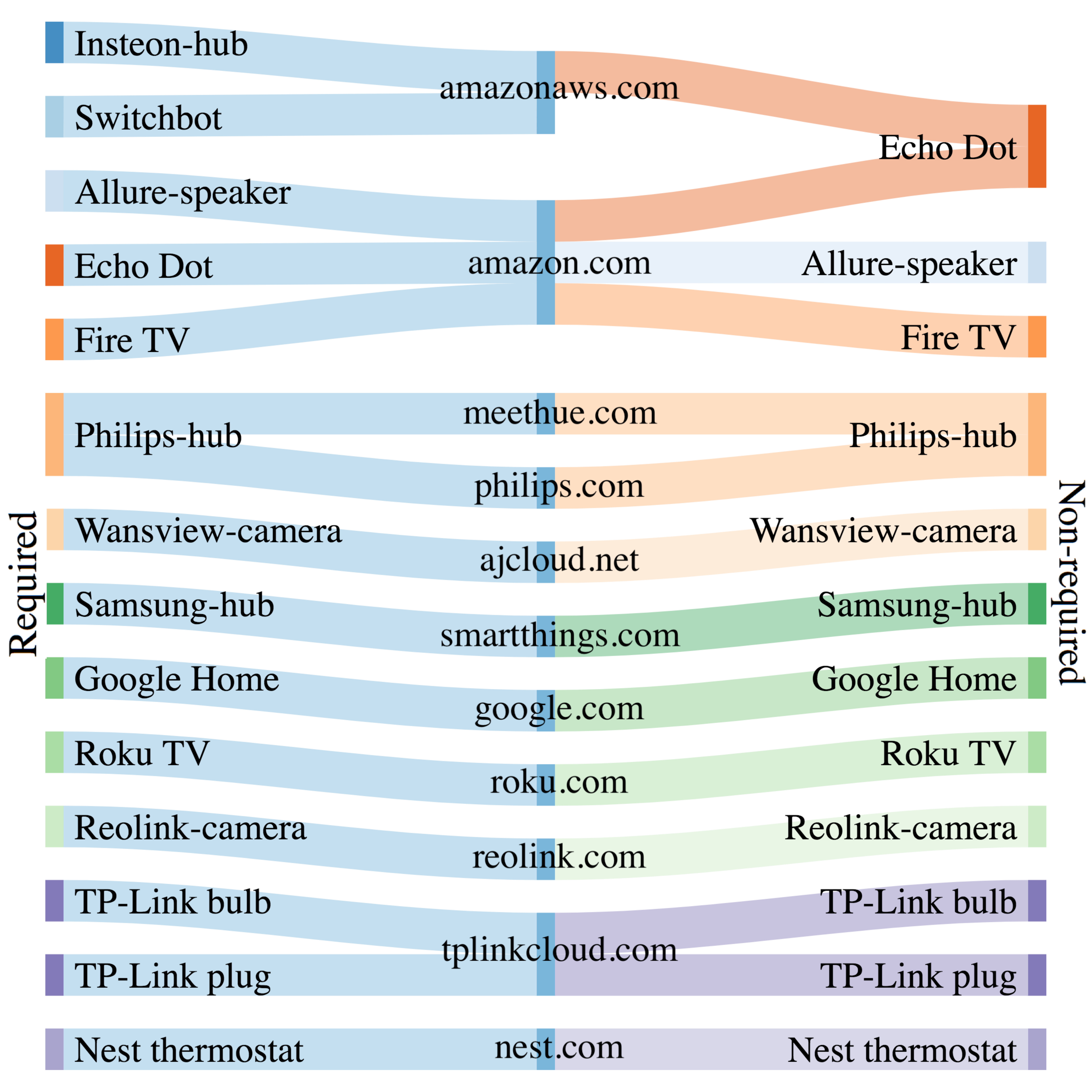}
		\caption{Second-level domain destinations that are both required and non-required. The width of the flow is proportional to the number of individual destinations contacted by each device. }
		\label{fig:sankey_sld}
	\end{center}
\end{figure}

\new{\textbf{Using only second-level domains (SLDs).} We now investigate whether SLDs are sufficient for identifying non-required destinations.  Fig.~\ref{fig:sankey_sld} shows the list of SLDs that are simultaneously required and non-required, and the list of devices contacting them as required (on the left) and non-required (on the right): for 12 devices contacting the 11 SLDs in the figure, SLDs are not specific enough since both essential and non-essential traffic use the same SLDs. For the remaining cases (19/31 devices), using only SLDs is sufficient. While this simplification of using SLDs is effective for identifying non-essential traffic for the majority of the devices we tested, it nonetheless would lead to mislabeling traffic for a significant fraction of devices (12/31). }



\noindent \new{\textbf{Using protocols and ports.} We determine whether the IP protocol and port are alone sufficient to detect traffic to non-required destinations.
The answer is generally no: we find that most non-required destinations consistently use HTTPS (TCP/443), with just the following exceptions:
two domains using HTTP (TCP/80) (\texttt{fireos\-captive\-portal\-.com} and \texttt{diagnostics\-.meethue\-.com}), and  ICMP packets sent to two IP addresses 
by the \textit{Bosiwo camera}.}

\new{ \subsection{Experiments with Additional Functions}
\label{sub:add_funct}
In the previous analyses, we consider, for each device, only the main function.
We now investigate if the list of required/non-required destinations changes when we consider \emph{additional functions}.
To this end, we increase functionality coverage by testing, in addition to the main function, at least three additional functions per device, listed per category in Table~\ref{table:functions}. We then recreate the lists of (non)-required destinations as follows: required destinations are the ones needed by \emph{at least} one tested function, while non-required destinations are the ones that are not needed by \emph{any} tested functions.}

\new{After testing additional functions, we see no changes in the list of non-required destinations for devices in all categories except for smart speakers: all of them contact several additional non-required destinations when asked for streaming music on YouTube/Amazon.
For example, the Google Home device contacts 4 additional non-required destinations, two of which are third parties (\texttt{googleadservices.com, googleads.g.doubleclick.net}).
With respect to required destinations, only five devices require new destinations to fulfill one of the additional functions: Yi camera (enable motion), Honeywell thermostat (adjust the temperature based on the weather), Google Home (stream music on YouTube), and Echo Dot/Allure (stream music on Amazon).
The lists of destinations contacted by additional functions in our tests are reported (in parenthesis) in Table~\ref{tab:reqnon-reqdestinations} in Appendix~\ref{sec:appendix}.}

\new{
\noindent\textbf{Takeaways.} While testing additional functions, non-required destinations are unchanged for 90.32\% of the devices, meaning that even if we only test the main function, we cover the vast majority or non-required destinations. 
Only five devices have one additional required destination, which is required and used only for one of the additional functions, suggesting that it is common for required destinations to fulfill more than one function.
While streaming content, smart speakers contact up to 4 non-required destinations and blocking those destinations does not break any additional functions.\footnote{Despite blocking advertisement destinations, there is no change in advertising behavior.}}

\new{

\begin{table}[t]
	\centering
		\resizebox{1.0\columnwidth}{!}{	
		\begin{tabular}{>{}l>{}r|>{}c|>{}c}
			\multirow{1}{*}{\textbf{Category}} &\multirow{1}{*}{\textbf{Additional Functions}} & \textbf{Req. \#}& \textbf{Non-Req. \#}\\			
\hline
\hline
Camera & Recording, get clip recordings, enable motion & 1 & 0\\	
\hline
\hline
Home-automation 	& Schedule, timer, set status, &&\\
				& set temperature, check water level & 1 & 0\\
\hline 
\hline
Hub & Schedule, timer, set status & 0 & 0\\
\hline
\hline
Speaker 		& Wikipedia search, google search,&& \\
				& play music on YouTube/Amazon & 5 & 6*\\
\hline
\hline
Video			& Sleeping mode, timer, add to watch list &  0 & 0 \\
\hline
\end{tabular}
}
\vspace{-0.5mm}
\caption{List of additional tested functions per category and number of additional required/non-required destinations. Only the additional function ``*playing music on YouTube/Amazon" triggers additional non-required destinations for smart speakers.} 
\label{table:functions}
\end{table}}

\subsection{Similarities with Existing Blocklists}
\label{sub:blocklists}

To conclude this section, we determine whether any of the observed required or non-required
destinations appear on blocklists from prior work. 
This can help clarify if any of such lists can be effective in also
blocking non-essential IoT traffic, or if they are likely to break some IoT functionality.
In this analysis we use the blocklists considered by Varmarken et. al~\cite{varmarken2020tv},
who evaluated the effectiveness of DNS-based blocklists to prevent smart TVs from accessing advertising and
tracking service domains.
In particular, they consider the most relevant blocklists to smart TVs \new{(actively managed)}, which are Pi-hole Default~\cite{Pi-hole-block}, the Firebog~\cite{TF}, Mother of all
Ad-Blocking (MoaAB)~\cite{MoaAB}, and StopAd~\cite{SATV}.

Table~\ref{table:block-lists} shows the devices having at
least one non-required destination. 
The table shows that existing blocklists contain very few of such destinations only for the most popular devices. 
Out of the \numdestnonreq{} non-required destinations, the most (\ie six) are obtained in the Firebog list. 
This is not surprising as the Firebog merges many popular blocklists into a single one.
The second most successful blocklist is the Pi-hole, which blocks four non-required destinations, 
the third is MoaAB blocking two, and the last is StopAd, which does
not contain any of the non-required destinations. 

Regarding the presence of required destinations in existing blocklists, we have not
found any, which means that current popular blocklists should not break the
functionality of the devices we tested.


\begin{table}[t]
	\centering
		\resizebox{1.0\columnwidth}{!}{	
		\begin{tabular}{m{0.01\columnwidth}p{0.252\columnwidth}||c|c|c|c|c}
			&\textbf{Device} & \textbf{Non-required \#} & \textbf{Pi-hole} & \textbf{Firebog} &
      \textbf{MoAB} & \textbf{StopAd} \\ \hline \hline
\multirow{5}{*}{\rotatebox[origin=c]{90}{\small Camera}} & Bosiwo  & 2 & 0 & 0 & 0 & 0 \\ 
\cline{2-7}
&iCSee     & 4 & 0 & 0 & 0 & 0 \\ 
\cline{2-7}
&Reolink  & 1 & 0 & 0 & 0 & 0 \\ 
\cline{2-7}
&Wansview   & 6 & 0 & 0 & 0 & 0 \\
\cline{2-7}
&Yi         & 2 & 0 & 0 & 0 & 0 \\ 
\hline
\hline
\multirow{4}{*}{\rotatebox[origin=c]{90}{\small Home-auto}}  & Nest thermostat   & 1 & 0 & 0 & 0 & 0 \\ 
\cline{2-7}
&TP-Link Bulb      & 3 & 0 & 0 & 0 & 0 \\ 
\cline{2-7}
&TP-Link Plug      & 3 & 0 & 0 & 0 & 0 \\ 
\cline{2-7}
&Xiaomi rice & 4 & 0 & 0 & 0 & 0 \\ 
\hline
\hline
\multirow{2}{*}{\rotatebox[origin=c]{90}{\small Hub}} & Philips & 2 & 0 & 0 & 0 & 0 \\ 
\cline{2-7}
&Samsung  & 1 & 0 & 0 & 0 & 0 \\ 
\hline
\hline
\multirow{3}{*}{\rotatebox[origin=c]{90}{\small Speaker}} & Allure   & 2 & 0 & 0 & 0 & 0 \\ 
\cline{2-7}
& Echo Dot          & 7 & 1 & 1 & 0 & 0 \\ 
\cline{2-7}
& Google Home       & 5 & 0 & 0 & 0 & 0 \\ 
\hline
\hline
\multirow{2}{*}{\rotatebox[origin=c]{90}{\small Video}} &Fire TV & 11 & 2 & 3 & 1 & 0 \\ 
\cline{2-7}
&Roku TV  & 8 & 1 & 2 & 1 & 0 \\ 
\hline
\hline
&\textit{Total}& \textit{62} & \textit{4} & \textit{6} & \textit{2}  & \textit{0} \\ 
\end{tabular}
}
\vspace{-0.5mm}
\caption{Similarity to existing blocklists. Comparison, by device, of the total number of non-required destinations with the number of such destinations that are present in various blocklists.}
\label{table:block-lists}
\end{table}

\noindent \textbf{Takeaways.} Most (91\%) destinations we have identified as
non-required do not appear on any of the existing blocklist, making them 
inadequate for mitigating non-essential traffic in the consumer IoT context.
\new{This occurs because existing blocklists primarily target websites and smart TVs, 
while we consider a broader range of IoT device categories that use different destinations.}

\section{Mitigating Non-essential Traffic}
\label{sec:design}

We answer our last research question by discussing how to limit IoT information exposure in practice.

\subsection{Blocking Strategies}

\noindent \textbf{Deny-listing:} 
\emph{blocking non-required destinations and allowing the rest of the traffic}.
This strategy only works under the assumption that non-required destinations are \emph{stable}, \ie they do not change over time.
To verify this, we measured the non-required destinations at several points in time over \nummonths{} months: May, July, and October 2020.
We then compared the three lists of non-required destinations and verified that there are no differences.
This means that the destinations that were non-required during our first set of experiments were \emph{still} contacted and non-required \nummonths{} months later. 

For this reason we consider all non-required destinations we encountered so far as stable.
Having a vast majority of stable non-required destinations means that a \deny blocking strategy is feasible because it does not need frequent updates on its blocklists, with low risk of allowing non-essential traffic and/or breaking the device functionality.
The drawback of this approach is that the possible appearance of new non-required destinations would not be mitigated.

\noindent \textbf{Allow-listing:}
\emph{allowing required destinations and blocking the rest}.
The assumption of this strategy is that required destinations do not change over time.
We also verified this assumption on the same three sets of experiments over \nummonths{} months, noticing that required destinations also do not change.

The stability of required destinations makes an \allow approach also feasible, without breaking the device functionality.
The advantage of this approach is that it has the highest mitigation potential, since existing and future non-required destinations will be blocked, but it also carries the highest risk of breaking the functionality of the device since if in the future a function requires a new destination, it will be blocked until the list of required destinations is updated.

\noindent \textbf{Choosing a blocking strategy.}
Based on the considerations above, choosing between a \deny and \allow blocking strategy depends on the priority between functionality and mitigation.
We believe that for the typical home IoT scenario a \deny strategy may be more appropriate, since maintaining functionality is a high priority (and mitigation of newly blockable destinations can be addressed through periodic blocklist updates).
In critical scenarios where privacy and security is a priority over functionality (\eg enterprise deployments), \allow may be the more appropriate.

\subsection{\new{Maintenance of Blocklists}}
\label{sub:maintenance}
\new{IoT systems may change the set of destinations they contact over time (\eg via firmware updates or server-side changes), potentially requiring updating the blocklists so they remain effective. 
While we did not observe such a change in \nummonths{} months, this may occur over longer periods.}
\new{Since all the steps of our approach are automated (except for the creation of probe and trigger scripts, which is manual, but only needed once), the measurement of (non-)required destinations can be easily iterated to keep the blocklists updated.
To minimize the risk of triggers/probes failing (\eg changes in the device interaction interface), we rely on our probe evaluation algorithm (see \S\ref{sub:probesevaluation}), which is run before measuring the destinations. Specifically, if a probe becomes inaccurate for a function or device, experiments are disabled and the problem is reported so that a human maintainer knows to update the affected trigger/probe scripts.
We anticipate that blocklists and the library of trigger/probe scripts will be maintained via options like crowdsourcing or via organizations that conduct our automated measurements on a regular basis (and share the outcomes), similar to what happens for blocklists for web/mobile-app trackers and advertisers.}


\subsection{Design of a Blocking System}

 \begin{figure}
  \centering
  \includegraphics[width=1.0\linewidth]{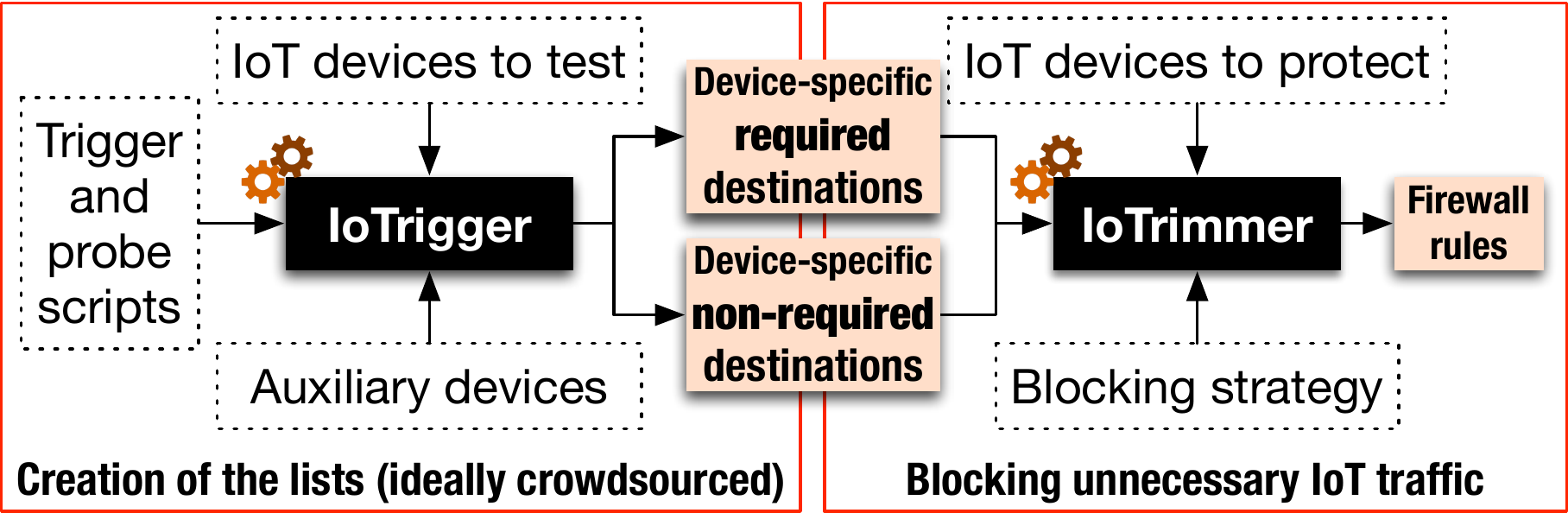}
  \caption{\new{Design of a blocking system. \trigger{} measures required/non-required destinations, while \blocker{} uses them to block non-essential IoT traffic by defining firewall rules. }}
  \label{fig:system}
\end{figure}

\new{To mitigate non-essential IoT traffic, we propose a blocking system composed of two
components: \trigger{} and \blocker{} (see Fig.~\ref{fig:system}). The former runs 
the methodology in~\S\ref{sec:method} to produce (non-)\-required
destination lists, and the latter uses such lists with a
blocking strategy to generate firewall traffic-blocking rules.}

\noindent \new{\textbf{\trigger{}}.
This component runs on a router providing connectivity to a set of IoT devices to test.
It manages the lifecycle of functionality experiments for each device, 
including the invocation of user-provided trigger and probe
scripts, and to finally produce (non-)required destination lists.
To work, \trigger{} needs the IoT devices connected to the same router,
the list of their IP addresses, the scripts to trigger and probe their functions,
and any other auxiliary devices (\eg devices used by trigger/probes scripts).
Given this, \trigger{} will run the experiment and
generate the destinations lists without any human interaction.
We implemented a command-line prototype of \trigger{}, which includes a library of 
probes and triggers scripts that support the IoT devices we tested.
Anyone owning the same IoT devices, and the proper trigger devices
(\eg Android phones) can use the \trigger{} prototype to reproduce our results.
For new devices and new functions, new trigger/probe scripts must
be added.} 

\noindent \new{\textbf{\blocker{}}.
This component runs on a router and uses the destination lists produced by \trigger{}
to determine which destinations to block.
\blocker{} takes as input these lists, the list 
of IoT devices (and their IP addresses) to be protected, and the blocking 
strategy (for generating firewall rules).
These rules, the final output of \blocker{}, are installed in the router to
block non-essential IoT traffic.
We implemented a prototype of \blocker{} (Appendix~\ref{sec:iotrimmerproto}). 
It comes preconfigured with the  \deny{}  blocking strategy and uses the blocklist of 
\numdestnonreq non-required destinations we found for our set of \numdev IoT devices. 
This \blocker{} prototype automatically detects devices connected to the IoT network: 
the user is provided with a web interface to associate the detected device
with the ones we have analyzed, so the prototype can automatically apply proper firewall rules.}

\subsection{Effectiveness of a Blocking System}

\noindent \textbf{\new{Effectiveness Evaluation.}}
\new{To measure the effectiveness of our blocking system in terms of preserving desired functionality, we run the following test for \numevaldays{} days on our \numdev{} IoT devices: we first protect the devices using \blocker{} with a \deny blocking strategy, then we use trigger and probe scripts to run their main functions once per day at different hours and check whether they work.
We found all of the \numfunct{} function invocations were successful and thus \blocker{} is effective at blocking without breaking for the devices we tested.}

\noindent \textbf{Risk of allowing non-essential traffic.}
Independently from the blocking strategy used, our approach tries to block non-essential traffic that is produced by non-required destinations. However, it is possible that some devices use (or may use in the
future, to elude our blocking strategy, \new{see \S\ref{sub:limitation})} the same destination for both essential and non-essential traffic.
In this case a future improvement of our approach is to look not just at the destination, but also at other traffic characteristics to find more distinguishing features, and
then filter the traffic based on such features.

\noindent \textbf{Risk of breaking device functionality.}
Our work is motivated by the fact that most consumer IoT devices are relatively simple, offering functions that are easy to test such as changing the state of a light, or asking a question to a smart speaker.
Since we test functionality similarly to how a generic algorithm is tested, in the case of complex functionality, we cannot prove that it works for every possible input (the total correctness of an algorithm is not decidable).
Due to this limitation we expect cases where complex functionality that has not been fully tested may break.
Unfortunately this is a limitation on most blocking systems (\eg ad-blockers~\cite{4939739} preventing a website from loading non-ad content correctly), which can be partially addressed by increasing the coverage of the functionality tested to the maximum possible extent, and refining blocklists accordingly.


\section{Discussion}
\label{sec:discussion}

We now discuss implications of our findings and limitations of our approach.

\subsection{Implications}
\noindent \textbf{Purpose of non-required destinations.} 
We know that non-required destinations are not essential for the main function of the device, but their intended purpose remains an open question,
particularly whether the purpose is benign or malicious.
One (optimistic) hypothesis is that they are required for other (\ie \emph{non-main}) functions that we did not test (\eg for syncing with a cloud service, checking for a firmware \new{update}).
A less optimistic hypothesis is that they are used for tracking purposes, given that some destinations (\eg \texttt{partnerad.l.doubleclick.net} and \texttt{netflix.com}) are third parties not related to the device manufacturer.

\noindent \textbf{Privacy considerations.}
In this work we have seen that many contacted destinations are not required for the device to operate.
The good news is that the \emph{quantity} (number of bytes) of data we have seen sent to the non-required destinations is very small compared to the rest of the traffic, as anything otherwise would be extremely suspicious. However, even if the amount of data is small, it is still a concern from a privacy
perspective, since it is still enough to signal the presence of a device and the
functions in use, as shown in previous work~\cite{Trimananda2020,
saidi2020haystack, Peek-a-Boo}.
Further, such non-required connections potentially violate the data minimization by design principle of some privacy regulations such as the GDPR~\cite{GDPR}.

A related question is whether device manufacturers reveal the purpose of these connections in their privacy policies. Unfortunately, many devices' privacy policies provide little information about how they use the data from their customers' devices~\cite{privacy_policy}. 
In many cases it is unclear whether a destination is used by the IoT device or the mobile app \new{controlling it}, and the behavior of some devices is not consistent with what is stated~\cite{subahi2018ensuring}.

\subsection{Limitations}
\label{sub:limitation}
\new{The experiments of this work have been executed on a fixed set of devices, and limited to \new{a subset of their functions}.}
We do not know if our results extend to other devices, after future firmware updates to existing devices, or for any additional functions.
However, our initial results are promising, suggesting that our methodology covers popular unmodified devices across different functional categories.
We expect that the non-essential traffic reported in this study represents a subset of all such traffic that our IoT devices generate.
As such, our findings represent a lower-bound of such traffic, using an approach that can be automated, \ie automatically detecting non-required destinations.

\noindent \textbf{Non-observable functionality.}
Our approach only works for device functions that can be tested using trigger and probe scripts.
\new{Some functions cannot be triggered (\eg device maintenance or synchronization tasks); to allow 
such functions as needed, one can periodically restart the device and unblock previously flagged non-required destinations temporarily 
to allow the maintenance connections to proceed.}

\noindent \textbf{\new{Firmware updates.}}
\new{While firmware updates are important for adding features and security patches to IoT devices, they may also introduce unwanted behavior~\cite{lgtvads}.
We believe it should be up to the user to decide whether to allow or block these updates.
By default our approach can block firmware updates if the corresponding destination(s) are not used for any essential function.
If a user chooses to allow firmware updates while blocking non-essential traffic, the following strategies may be used.
For unattended updates, we can use the ``non-observable functionality'' approach (\eg restart the device while keeping destinations unblocked for a set period of time).
For user-initiated updates, we can treat them as a device function, and use a dedicated set of trigger/probe scripts to detect what destinations are used by the firmware update function. 
Another approach is to allow traffic matching patterns that reveal the firmware update intention (\eg destinations containing strings such as ``fwupdate''). }


\noindent \new{\textbf{Scalability.}
Every step of our approach is fully automated, including the execution of probe and trigger scripts, with each function to be tested taking an average of 4 minutes and easily repeatable to allow frequent crowdsourced updates. 
However, the \emph{creation} of such scripts is a manual process that has to be repeated once for every function tested for each device.
A mitigating factor is that devices belonging to similar categories may reuse existing scripts with little modifications (\eg a simple change of tap coordinates for companion app triggers, and of screenshots for companion app probes).
Although not observed in the \nummonths{} months of our study, it is possible that trigger and probe scripts stop working and need to be manually modified when device functionality changes substantially (\eg via firmware updates).
We can identify such cases by periodically running our automated probe evaluation algorithm (see \S\ref{sub:probesevaluation}).}

\noindent \textbf{\new{Blocking granularity.}}
\new{Our approach focuses on destination-based blocking to reduce information exposure; however, other factors may be used to identify traffic that should be blocked, \eg time of day, traffic volumes, device-to-device communication.
One advantage of only considering destinations is that it is easy to automatically measure, and easy to block using simple firewall rules, without the need of fine-grained enforcement mechanisms that may not be readily deployable and may incur heavy overhead at runtime.}

\noindent \textbf{\new{Evading blocklists.}}
\new{To evade the blocklists a device can disable its functionality when any of its destinations are unreachable.
This limitation also exists in anti-tracking browser plugins, where a website is not allowed to load until anti-tracking software is disabled~\cite{nithyanand2016adblocking}. 
There is no simple defense against this evasion technique, but our approach can still block any non-required destinations where a device 
does not try to evade blocking, \eg destinations used by third-party apps.}

\noindent \new{\textbf{Third-party apps.}
Some devices include pre-installed third-party apps (\eg Netflix on video devices). 
In such cases, background traffic or content previewed on a menu screen may be considered required 
or non-required depending on whether the device owner wishes to use those apps. 
A limitation of our work is that we cannot know whether to block or allow the  
traffic for third-party apps without user input about which apps are required to work. 
As an example, we found that \texttt{netflix.com} was identified 
as non-required by default in our approach because it is not necessary for the menu screen to work. 
For users that subscribe to Netflix, we can include results from testing the Netflix app on the device 
and treat corresponding destinations as required.}


\noindent \textbf{Working with MUD profiles.}
Manufacturer Usage Description (MUD) profiles~\cite{mud-rfc,MUD-sigcomm2018w} allow manufacturers to declare the behavior of their devices (including the destinations contacted).
None of the devices we tested implements MUD profiles.
However, even if a destination is declared in the future, a MUD profile does not
help to determine if such destination is used for essential traffic only.
Hence, our approach is orthogonal to MUD profile enforcement 
and can work side-by-side with it.

\section{Related Work}
\label{sec:relatedwork}

Recent research has produced a number of tools to protect against undesirable
IoT traffic.  Haar and Buchmann presented FANE~\cite{FANE}, a firewall that
isolates IoT devices into a separate network. FANE allows communication only
with the learned set of IPs.  If a device contacts a new IP address, the user is
alarmed. FANE does not support blocking destinations based on domain names.
Simpson \etal~\cite{Simpson2017} focus on protecting IoT devices against known
vulnerabilities and automatically blocking traffic when a threat is identified.
Gupta \etal~\cite{Firewall4IoT} propose a firewall based on a Raspberry Pi with
simple \emph{iptables} rules to protect the devices from potential attacks.
Heimdall~\cite{Heimdall} focuses on protecting devices against hacks from
the Internet using a pre-learned \emph{allow-list}.  Lastdrager \etal~\cite{spin}
describe SPIN, a software tool for visualizing and blocking traffic
from IoT devices. 
None of these solutions focus on mitigating information
exposure nor blocking connections without breaking device functionality. 

Numerous commercial tools provide solutions to protect networks with IoT
devices, \eg ShieldIOT~\cite{shieldiot}, Fing~\cite{fing}, and
Bitdefender~\cite{bitdefender}. These approaches either rely on cloud-based
analysis of network traffic, target device manufacturers rather than
device users, block or allow the device as a whole, monitor the overall
amount of IoT traffic generated, or protect against known vulnerabilities
and attacks from the Internet. \blocker allows fine-grained control over
destinations contacted by the devices and protects users privacy by blocking the
unnecessary traffic generated by IoT devices. 

There are a number of existing tools for IoT privacy risk analysis.  For
example, IoT Inspector~\cite{huang2019iot} collects smart home traffic using ARP
spoofing. However, this tool focuses on the collection of data, rather than its
analysis or the blocking of non-essential traffic.  A recent
study~\cite{moniotr} of 81 consumer IoT devices shows that many IoT devices
expose information to first, support, and third parties. Additional research
uses traffic generated by the IoT devices to identify devices or
device activities~\cite{Peek-a-Boo, Apthorpe2016, Tahaei2020, Meidan2017,
Miettinen2017, Hafeez2020, Trimananda2020, saidi2020haystack}. 
Because \blocker can reduce the number of destinations contacted by IoT
devices, it reduces the attack surface and can prevent an eavesdropper to
identify users device or activity. 

Two recent IoT papers focus on strategies for defending user privacy against
potential eavesdroppers (\eg ISP). Apthorpe \etal~\cite{Apthorpe2019} propose
generating additional dummy network traffic that hides genuine IoT device
network traffic patterns from an observer, and Alshehri
\etal~\cite{Alshehri2020} proposes a similar approach using uniform random
noise. In contrast, our approach focuses on protecting users' privacy from
legitimate destinations that the IoT devices communicate with, not against
potential eavesdroppers. However, our approach can be integrated with the above
work to enhance user privacy against potential eavesdroppers. 


\section{Conclusion}
\label{sec:conclusion}

This paper demonstrated that it is feasible and effective to block non-essential network traffic from IoT devices, thus limiting the information they expose to other parties without breaking device functionality. We developed the first comprehensive method to automatically identify non-required destinations from network traffic, and analyzed the results of the corresponding experiments.

We found that \numdevblockable{} of the \numdev{} consumer IoT devices in our study contact destinations that are not required to fulfill \new{their main functions.} 
Most destinations (\numdestnonreq{} out of \numdest{}) are responsible for non-essential traffic, and such non-required destinations are relatively long-lasting in our study---
they did not change at all over \nummonths{} months.
The vast majority (91\%) of destinations responsible for non-essential traffic are not listed in any other \new{general} blocklist, demonstrating the benefits of our \new{device-dependent} approach.
\new{Finally, we produced a set of guidelines and a prototype of a blocking system to mitigate non-essential IoT traffic.}

To support further research, all software and data we produced as part of this work are publicly available at \url{http://iotrim.net/}.

\section{Acknowledgments}

\new{We thank the anonymous reviewers for their constructive feedback. The research in this paper was partially supported by the EPSRC (Databox EP/N028260/1, DADA EP/R03351X/1, HDI EP/R045178/1, and Impact Acceleration Account (IAA)), NSF (BehavIoT CNS-1909020, ProperData SaTC-1955227) and Consumer Reports (Digital Lab Fellowship for Daniel J. Dubois).}


\bibliographystyle{unsrt}
\bibliography{main}

\appendix


\section{\new{Grouping Ephemeral Destinations}}
\label{sec:grouping}

\new{During our destination-observing experiments (see \S\ref{sub:list}), some devices contact destinations that appear in less than 80\% of the experiment iterations.
We refer to such destinations as \emph{ephemeral destinations}.
To facilitate analysis and streamline blocklists, we developed two algorithms to automatically group ephemeral destinations into specific groups that cover ephemeral destinations in at least 80\% of the iterations. One algorithm is for ephemeral hostname destinations and the other for ephemeral IP destinations.}

\noindent \textbf{\new{Grouping hostname destinations.}}
\new{When a hostname is ephemeral (\ie it appears in less than 80\% of the iterations), we remove the first character from the domain name and replace it with a wildcard matching any number of characters (zero or more).
If the resulting group matches domains in at least 80\% of the iterations, we consider such group as a new hostname destination (and remove all the matching hostnames from the list of destinations).
If not, we repeat the process recursively by replacing additional characters with the wildcard up to the entire second-level domain.
For example, ephemeral domains \texttt{1.yy.com} and \texttt{2.yy.com} are replaced by the group \texttt{*.yy.com}, which also happens to be the entire second-level domain.}

\new{Note that our algorithm is also capable of finding groups that are \emph{more specific} than second-level domains.
For example, ephemeral domains \texttt{a-b-c.ww.com} and \texttt{b-b-c.ww.com} are replaced by the group \texttt{*-b-c.ww.com}, which is more specific than a second-level domain.}

\new{Across all our destination-observing experiments, we have found three groups matching ephemeral hostname destinations (\texttt{*.backblaze.com}, \texttt{*.cloudfront.net}, \texttt{*.googlevideo.com}), which also happen to be second-level domain names, since there were no more specific alternatives.
We have found no cases of ephemeral hostname destinations that could not be matches by one of the three groups above.}

\noindent \textbf{\new{Grouping IP destinations.}}
\new{When an IP address is ephemeral (\ie it appears in less than 80\% of the iterations), we perform a WHOIS query to get the IP mask that includes such IP address. 
If such IP mask matches an ephemeral IP in at least 80\% of the iterations, we consider such IP mask as a destination (and remove all the matching IPs from the list of destinations).
For example, if we have two ephemeral IPs 1.2.3.4 and 1.2.4.5, and for both of them we obtain a matching mask from WHOIS that is 1.2.0.0/16, we would use 1.2.0.0/16 as the grouped destination.}

\new{Across all our destination-observing experiments, we have found no cases of ephemeral IP addresses, and therefore we have no IP destination groups.
Still, should that happen in the future, this algorithm would be able to deal with such cases.}

\clearpage

\section{List of Required and Non-required Destinations}
\label{sec:appendix}

In this appendix we report, for each device, the list of required and non-required destinations. This is the data we used to produce part of the analyses in \S\ref{sec:destinations}.

From Table~\ref{tab:reqnon-reqdestinations}, we can confirm that \numdevblockable{} out of \numdev{} \newpage \noindent  devices contact at least one non-required destination.
In general, the number of non-required destinations tend to be larger than the number of required ones.

We believe this information, together with the \trigger and the \blocker software available at \url{http://iotrim.net/}, to be valuable for researchers, device manufacturers, and regulators to support and reproduce our findings.

\hspace{-1.13\linewidth}
\begin{minipage}{1\textwidth}

\begin{table}[H]
	\resizebox{1\textwidth}{!}{
		\centering
		\begin{tabular}{m{0.01\textwidth}p{0.15\textwidth}||p{0.07\textwidth}|p{0.5\textwidth}|p{0.5\textwidth}}
		\centering
			&\textbf{Device} & \textbf{Dest. \#} & \textbf{List of Required Destinations} & \textbf{List of Non-Required Destinations}\\ 
			\hline
			\hline
\multirow{9}{*}{\rotatebox[origin=c]{90}{\small Camera}} &\hspace{1mm} Blink & \centering 2 &rest-hw-prde.immedia-semi.com, cs-prde.immedia-semi.com  & \\
\cline{2-5}
&\hspace{1mm} Bosiwo & \centering 4 & \color{gray}\textit{145.239.253.48, 37.187.159.39}  & \color{gray}\textit{54.157.82.107}, \color{red}\underline{210.72.145.44}\\
\cline{2-5}
&\hspace{1mm} iCSee & \centering6   & \color{gray}\textit{47.91.198.64, 47.91.207.52} & \color{gray}\textit{47.52.222.172, 47.52.32.118, api.gdxp.com, oss-us-west-1.aliyuncs.com}  \\
\cline{2-5}
&\hspace{1mm} Reolink & \centering2    & p2p.reolink.com &pushx.reolink.com    \\
\cline{2-5}
&\hspace{1mm} Wansview & \centering 9  & \color{gray}\textit{cam-gw-isc-eu02.ajcloud.net, cam-tunnel-isc-eu02.ajcloud.net, fw-isc.ajcloud.net} &  \color{gray}\textit{159.65.95.225, 3.122.229.130 , ajcloud.net, htpdate.ajcloud.net, sdc-isc.ajcloud.net, *.backblaze.com}   \\
\cline{2-5}
&\hspace{1mm} Yi & \centering5 (1)  &  \color{gray}\textit{47.74.255.9, 47.88.59.209, 47.90.240.160}, (motiondetection-eu.oss-eu-central-1.aliyuncs.com) & api.eu.xiaoyi.com, log.eu.xiaoyi.com   \\
\hline
\hline
\multirow{18}{*}{\rotatebox[origin=c]{90}{\small Home-automation}} & \hspace{1mm} App Kettle & \centering2  & ak.myappkettle.com, query.jingxuncloud.com  &  \\
\cline{2-5}
& \hspace{1mm} Honeywell therm. & \centering2 (1)  & \color{gray}\textit{ihsu-prod-bl-003.cloudapp.net,lcc-prodsf-fwu.eastus.cloudapp.azure.com, (weather.clouddevice.io)}& \\
\cline{2-5}
& \hspace{1mm} Magichome & \centering1   & ra8816us.magichue.net& \\
\cline{2-5}
& \hspace{1mm} Meross opener & \centering1   & iot.meross.com& \\
\cline{2-5}
& \hspace{1mm} Nest thermostat & \centering3   &transport.home.nest.com, logsink.devices.nest.com  & frontdoor.nest.com  \\
\cline{2-5}
& \hspace{1mm} Netatmo weather & \centering1   & netcom.netatmo.net& \\
\cline{2-5}
& \hspace{1mm}  Smarter coffee  & \centering1   & prd19a.boxen.electricimp.com& \\
\cline{2-5}
& \hspace{1mm}  Smartlife bulb  & \centering1   & a.tuyaeu.com& \\
\cline{2-5}
& \hspace{1mm}  Smartlife remote  & \centering1   & a.tuyaeu.com& \\
\cline{2-5}
& \hspace{1mm}  Sousvide   & \centering1   & pc.anovaculinary.com& \\
\cline{2-5}
& \hspace{1mm}  Switchbot   & \centering1   & \color{gray}\textit{a2alhn2dfztqv9.iot.us-east-1.amazonaws.com}& \\
\cline{2-5}
& \hspace{1mm} TP-Link bulb & \centering4 & n-devs.tplinkcloud.com & euw1-api.tplinkra.com, n-deventry.tplinkcloud.com, use1-api.tplinkra.com \\
\cline{2-5}
& \hspace{1mm} TP-Link plug &\centering 4 & n-devs.tplinkcloud.com   & euw1-api.tplinkra.com, n-deventry.tplinkcloud.com, use1-api.tplinkra.com \\
\cline{2-5}
& \hspace{1mm}  Wemo plug  & \centering2   & \color{gray}\textit{api.xbcs.net, nat.xbcs.net}& \\
\cline{2-5}
& \hspace{1mm} Xiaomi rice-cooker& \centering7   & mi.com, ot.io.mi.com, \color{gray}\textit{120.92.65.243}  & \color{gray}\textit{183.84.5.203, 58.83.160.36, 123.125.102.215, 110.43.0.83}   \\
\hline
\hline
\multirow{5}{*}{\rotatebox[origin=c]{90}{\small Hub}} & \hspace{1mm} Insteon & \centering1   &  \color{gray}\textit{lb-connect-insteon-com-503033429.us-east-1.elb.amazonaws.com} &  \\
\cline{2-5}
& \hspace{1mm} Lightify & \centering3   & \color{gray}\textit{srm-emea-p01-lb02.arrayent.com, 35.157.95.104, 35.159.20.196} &  \\
\cline{2-5}
& \hspace{1mm} Philips & \centering4   &  dcp.dc1.philips.com, ws.meethue.com  & diagnostics.meethue.com, ecdinterface.philips.com \\
\cline{2-5}
& \hspace{1mm} Samsung & \centering3    & api.smartthings.com, dc.CoNnect.SMaRTThInGs.cOm  & fw-update2.smartthings.com \\
\cline{2-5}
& \hspace{1mm} Sengled & \centering2   & eu.cloud.sengled.com, \color{gray}\textit{18.195.119.104} &  \\
\hline
\hline
\multirow{10}{*}{\rotatebox[origin=c]{90}{\small Speaker}} & \hspace{1mm} Allure& \centering3 (3)  &   bob-dispatch-prod-eu.amazon.com, (m.media-amazon.com, tinytts.amazon.com) &   \color{gray}\textit{api.amazon.com, d1enchupjctwud.cloudfront.net, (msh.amazon.com)} \\
\cline{2-5}
& \hspace{1mm} Echo Dot & \centering10 (3)  &   api.amazon.com, bob-dispatch-prod-eu.amazon.com,  unagi.amazon.com, (m.media-amazon.com, tinytts.amazon.com)  &arcus-uswest.amazon.com, *.cloudfront.net, device-metrics-us.amazon.com, dp-gw.amazon.com, fireoscaptiveportal.com, \color{gray}\textit{prod.amcs-tachyon.com} \color{black}, s3-1-w.amazonaws.com, (msh.amazon.com) \\
\cline{2-5}
& \hspace{1mm} Google Home & \centering9 (5)    &connectivitycheck.gstatic.com, home-devices.googleapis.com, play.googleapis.com, www.google.com, (*knez.googlevideo.com) & youtube-ui.l.google.com, clientservices.googleapis.com, fcm.googleapis.com, *.googlevideo.com, storage.googleapis.com, (tools.google.com, www.youtube.com, \color{red}\underline{www.googleadservices.com, googleads.g.doubleclick.net)}\\
\hline
\hline
\multirow{10}{*}{\rotatebox[origin=c]{90}{\small Video}} & \hspace{1mm} Fire TV & \centering14  & api.amazon.com, unagi-eu.amazon.com, \color{gray}\textit{youtube.com} &\color{red}\underline{aax-eu.amazon-adsystem.com}, \color{black}arcus-uswest.amazon.com, bob-dispatch-prod-eu.amazon.com, *.cloudfront.net, device-metrics-us.amazon.com, api.amazon.com, ktpx-eu.amazon.com, \color{red}\underline{api-global.eu-west-1.prodaa.netflix.com}\color{black}, mas-ext-eu.amazon.com, mas-sdk.amazon.com, msh.amazon.com\\
\cline{2-5}
& \hspace{1mm} Roku TV& \centering10    & api.sr.roku.com, \color{gray}\textit{youtube.com} &    \color{red}\underline{api-global.eu-west-1.prodaa.netflix.com}\color{black}, configsvc.cs.roku.com, cooper.logs.roku.com, \color{red}\underline{customerevents.eu-west-1.prodaa.netflix.com}, \color{red}\underline{ichnaea.eu-west-1.prodaa.netflix.com}, \color{red}\underline{partnerad.l.doubleclick.net}\color{black}, scribe.logs.roku.com, \color{red}\underline{uiboot.eu-west-1.prodaa.netflix.com}\\
\hline
\hline
\textit{Total} & \centering  \textit{31} &\centering  \textit{119 (13)} &\textit{57 (7)} &  \textit{62 (6)}  \\ 
		\end{tabular}
	}
	\vspace{-0.5mm}
	\captionsetup{width=0.99\textwidth}
	\caption{Required and non-required destinations per device. {Colors identify the destination party type (see~\S\ref{sub:deviceparty})}: first party, \color{gray}\textit{support party}\color{black}, and \color{red}\underline{third party}\color{black}. In parenthesis the additional destinations for the additional functions. Only ``playing music on YouTube/Amazon" triggers additional non-required destinations for smart speakers.} 
	\label{tab:reqnon-reqdestinations}
\end{table}

\end{minipage}

\clearpage

\section{\blocker Prototype}
\label{sec:iotrimmerproto}

\begin{figure}[H]
  \begin{center}
  \includegraphics[width=1\textwidth]{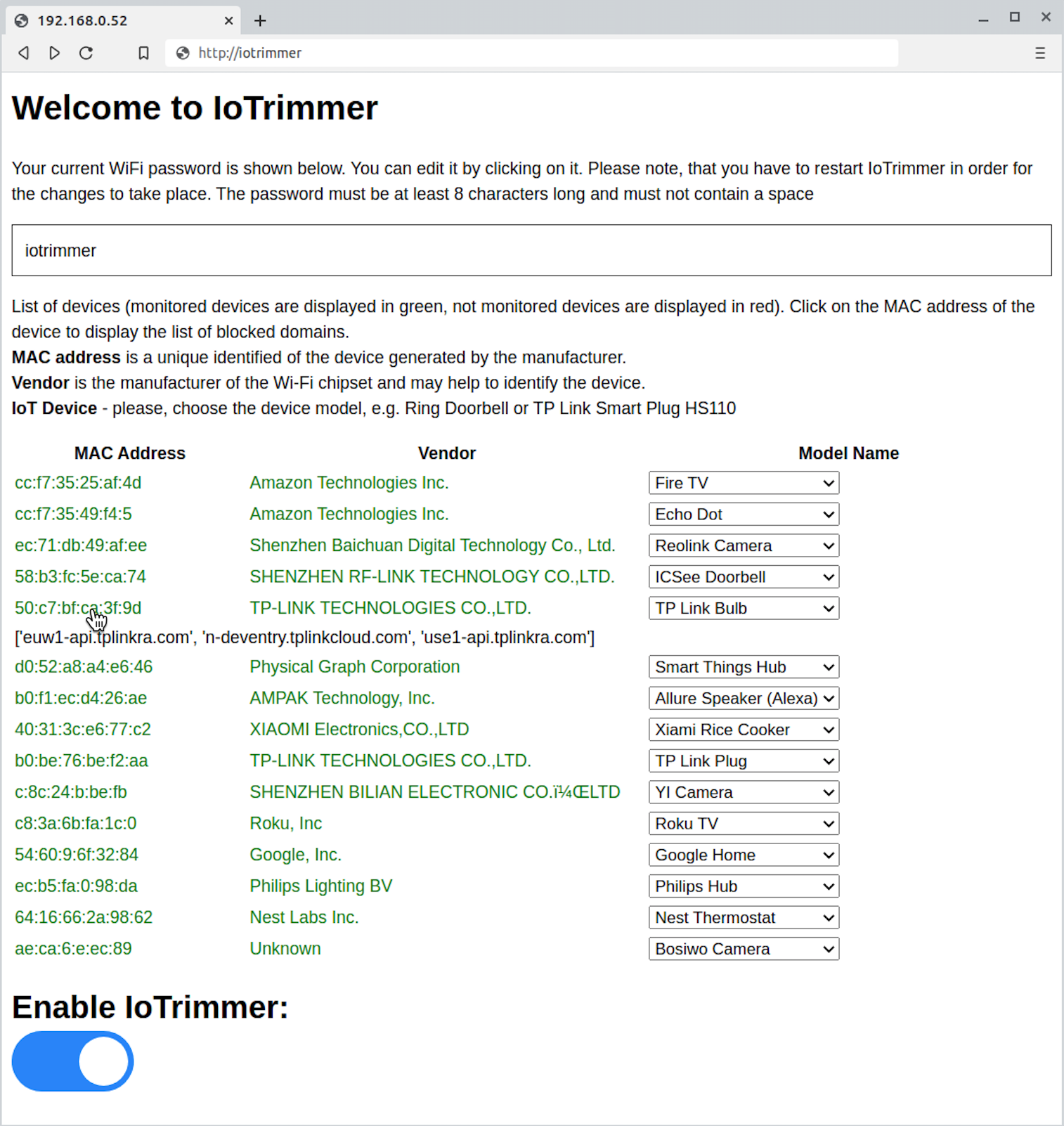}
  \caption{Prototype implementation of \blocker.}
  \label{fig:prototype}
  \end{center}
\end{figure}

We have implemented a prototype version of \blocker{}. Fig.~\ref{fig:prototype}
shows its web interface. When a new device is connected to \blocker{} its
MAC address appears on the list. 

\newpage

The user then chooses which device is connected
to \blocker{}. 
The blocklist (\iolist) is regularly updated from the Internet and
automatically applied to all connected devices. Users can click on a device to
display the list of blocked destinations. 

\end{document}